\documentclass[12pt]{article}
\usepackage[dvips]{graphicx,psfrag}

\def\beqa{\begin{eqnarray}}
\def\eeqa{\end{eqnarray}}
\def\papertitlepage{\baselineskip 3.5ex \thispagestyle{empty}}
\def\Title#1{\vspace{1.5cm}\begin{center}
 {\Large\bf #1} \end{center}
\vspace{1cm}}
\def\Authors#1{\begin{center} {\large #1} \end{center}}
\def\Abstract{\vspace{0.3cm}\begin{center} {\large\bf Abstract}
           \end{center} 
}
\begin{document}
\papertitlepage
\vspace*{-1 cm}
\hfill 
\begin{minipage}{4cm} {UT-Komaba/03-8}
          \par\noindent  {April  2003}\par\noindent {hep-th/0304237}  
\end{minipage}
\Title{D-brane Dynamics and Creations of \\
Open and Closed Strings after Recombination}

\Authors{{   Takeshi Sato
\footnote{tsato@hep1.c.u-tokyo.ac.jp}} \\
 \vskip 3ex
{\itshape  Institute of Physics, University of Tokyo, \\
3-8-1 Komaba, Meguro-ku, Tokyo 153-8902 Japan} \\
}


\Abstract

A quantum-mechanical technique is used within the framework of U(2) 
super-Yang-Mills theory to investigate what happens in the process
after recombination of two D-p-branes at one angle.   
Two types of initial conditions are considered, 
one of which with $p=4$ is a candidate of inflation mechanism.
It is observed that the branes' shapes come to have three extremes 
due to localization of tachyon condensation. 
Furthermore, open string pairs connecting 
the decaying D-branes are shown to be created;
most part of the released energy is used to create them.
It also strongly suggests 
that creation of closed strings happens afterward.
Closed strings as gravitational radiation from the D-branes are also 
shown to be created. 
A few speculations are also given on implications of the above phenomena 
for an inflation model.

\newpage

\section{Introduction}
\setcounter{equation}{0}

Recombination of D-branes at angles are processes 
increasing its value recently 
from both phenomenological 
and theoretical point of view, related to the presence of
tachyonic modes which appear as modes of open strings
between the two D-branes\cite{angle}.  
In the brane inflation scenarios,
these are some of the promising candidates 
to explain the mechanism of inflation\cite{halyo}\cite{hirano1}%
\cite{ky}\cite{oneangle}\cite{int2}\cite{hirano2}\cite{tye1}%
\cite{2angles}\cite{bro}\cite{br1}.
In addition, among various string constructions of Standard Model,
intersecting brane models are one class of hopeful
candidates\cite{bl1}\cite{chiral}\cite{bl2}\cite{getjust}\cite{higgs}%
\cite{koko1}\footnote{
For a recent review and other references, see e.g.
ref.\cite{ur1}.},
in some of which it has been proposed that
the recombination process occurs as Higgs mechanism%
\cite{chiral}\cite{higgs}\cite{koko1} (see also \cite{bachas}\cite{koko2}). 
On the other hand, this system 
and process can be regarded as a generalized 
setting of $D\bar{D}$ system and its annihilation process, 
which has been studied
thoroughly\cite{sen1}\cite{y1}\cite{y2}\cite{sen5}%
\cite{lar}\cite{tera}\cite{tye2}. 
In this way, this is one of the most important phenomena 
to explore at present in string theory.

The behavior of the system after  recombination, however, 
had been almost unexamined until recently.
Suppose one consider a recombination of two D-branes at one angle, 
wrapping around some cycles in a compact space.
One might expect that after the recombination, 
each brane would begin to take a ``shorter cut'' and then make 
a damped oscillation around some stable configuration of branes,
while radiating RR gauge and gravitational waves (and some others),
leading to the stable configuration.
However, the case was that only the shape of the recombined branes 
at the initial stage was merely inferred\cite{pol}, 
though the final (infinitely late time) 
configuration can be determined in some cases (e.g. see ref.\cite{2angles}).

Recently, in ref.\cite{hn}, K. Hashimoto and Nagaoka
proposed that a T-dual of super Yang-Mills theory (SYM)
can be an adequate framework
to describe the process, and in the case of two D-strings at one angle and via 
classical analysis, they presumed how the shapes of the branes develop 
at the very initial stage\footnote{Just after ref.\cite{hn}, the process 
is also 
analyzed via tachyon effective field theory in ref.\cite{wh} and very recently,
the subsequent work was done by K. Hashimoto and W. Taylor in ref.\cite{khw}.}

Main purpose of this paper is to investigate 
the process after
the recombination via SYM
``more rigorously'' and to reveal what happens in 
the process.
To be concrete, we will focus on the two points: 
time-evolution of the D-branes' shape,
and behavior or creations of fundamental strings.
The former can be of value in that 
we describes time-dependent, i.e. 
dynamical behavior of curved D-branes though for 
a rather short time-scale (which enables us to discuss
e.g. gravitational waves radiated by the branes).  
The former might also be regarded as a first step toward the
understanding of branes' shapes after the recombination 
in intersecting brane models as Standard Model,
which are responsible for physical quantities after the condensation
of Higgs fields. 
The latter point is what happens in the process itself.   
What we mean by ``more rigorously'' is that 
we set concrete initial conditions and make a quantum-mechanical 
analysis:
These are necessary steps in order to understand precise behavior of 
the system, because the 
system, its tachyon sector in particular, 
is essentially a collection of inverse harmonic oscillators,
and its behavior depends crucially on
the initial fluctuations of the system,
which can only be evaluated appropriately via a quantum analysis,
but has not been done at least in 
this setting.\footnote{
In the case of inflation and other systems, see 
ref.\cite{guth}\cite{boya1}, and for a non-perturbative analysis,
see ref.\cite{linde1}.}

The second purpose of this paper is to discuss
implications of the above two points (phenomena),
for the inflation scenario of the setting that two D-4-branes are
approaching each other at one angle as in
ref.\cite{oneangle}\cite{tye1},
aiming at the understanding of the whole process of the scenario.

For the above two purposes,
we consider two types of initial conditions for  
branes. As we will discuss later,
in order to make a quantum-mechanical analysis,
we need the fact that all the modes (except U(1) gauge fields on each of
the branes) 
have positive frequency-squareds at
$t=0$. So,
the first one we consider (denoted as case (I)) is that 
two D-p-branes have been parallel until $t=0$,
but are put
intersected at one angle $\theta$ at the instant $t=0$.
The second one (the case (II)) is that one of parallel two
D-p-branes were rotated 
by a small angle $\theta$ and are approaching each other very slowly.
The case (I) is one of the simplest conditions, rather easier 
to study the process itself.
The case (II) is more practical;
the case with $p=4$ is one of the hopeful setting for 
the brane inflation scenarios\cite{oneangle}\cite{tye1}.
From a practical viewpoint, the case (I) may seem less realistic,
but it may also be able to be regarded as some local or rough
approximation to more complicated setting of D-branes.

The outline of this paper is summarized as follows:

We consider two D-p-branes for each initial condition and for each 
$p\ge 1$, within the framework of a T-dual of SYM, 
a low energy effective theory of open strings around the D-branes
when distances between the branes are small, as in ref.\cite{hw}\cite{hn}.
Detailed set-ups are given in section 2.

In section 2 and 3 we evaluate time-evolution of typical amplitudes of 
tachyonic fluctuations in the following way; 
we consider fluctuations around the D-branes 
and make mode-expansions of them to examine their spectra.
Then, defining typical amplitudes (absolute values) of the fluctuations
in terms of VEV's of their absolute square,
we obtain them by using a time-dependent wave function of
each mode and focusing on contribution of tachyonic modes.
We note that though we use WKB approximation (second order one 
in fluctuations), 
it is enough good in the cases until the tachyons blow up,
because higher order terms are suppressed by a factor of
$O(g_{s})$, as we will discuss later.

In section 4 we discuss time-evolution of  D-brane's shape after 
recombination; we show that a seemingly queer behavior of the shape 
appears and explain why it happens. 
The information of the shape is obtained from a  
transverse U(2) scalar field 
by diagonalizing  its VEV's (i.e. choosing a certain gauge)
and looking at its diagonal elements\cite{hn}.\footnote{
For case (II), there appears a non-commutativity
between VEV's of scalar fields, which implies their uncertainty
relation. We will discuss this point in section 8.}
We denote this as ``physical gauge'' because in the gauge direct 
correspondence holds between d.o.f. of each open string starting from and 
ending on certain branes
and d.o.f. of each U(2) matrix element.

Then, in section 5 we argue that open string pairs (composed of strings
with opposite orientations) connecting the two decaying branes are
created. 
Our argument is based on the behavior of the U(2) gauge and scalar fields 
{\it in the physical gauge};
investigation of
its non-vanishing energy density, the correspondence of d.o.f. stated
above, and charge densities of two U(1) symmetries on the D-branes
lead to such conclusion.
We also evaluate number density of the open strings and 
distribution of energy to the open strings, and show that 
most part of the released energy are used to create the pairs.

In section 6, we discuss creation of closed strings, going out of the
framework of SYM: We show that the decaying branes radiate  
closed strings as gravitational wave, regarding the branes as a
collection of massive small objects; its energy flux is evaluated
quantitatively. 
We also discuss the system after some relatively long time, i.e. 
beyond the reach of SYM;
creation of closed strings 
are strongly suggested after the two branes are separated further.

In section 7, we discuss implications of the above results for 
a brane inflation scenario, on the efficiency of the reheating and 
mechanism to create ``reheated'' gauge fields and fermions.

Finally in section 8 we present summary and discussion;
we discuss some points on 
non-commutative scalar fields appearing in the case (II).
In the appendix we give a short review of giving
a width of a Gaussian wave function for each mode at arbitrary time.


\section{Set-ups and preliminaries}
\setcounter{equation}{0}
\setcounter{footnote}{0}

In this section we present set-ups and preliminaries for a
quantum-mechanical analysis for the case (I) and (II).

First we present our notation and set-ups;
in this paper we set $l_{s}= 1 $ for convenience 
(and revive it when needed), 
and set the string coupling $g_{s}$ very small 
and consider D-p-branes in a 10-dimensional
flat target space with a metric $g_{\mu\nu}={\rm diag}(-1,1,..,1)$
and coordinates $x_{M}$ ($M=0,..,9$). The world-volumes are parametrized by 
$x_{0},x_{1},\cdots x_{p}$.
We compactify space dimensions $x_{p},\cdots,x_{9}$ 
on a (10-p)-torus with periods $L_{a}$ for $a=p,\cdots,9$.
We denote $x_{0}$ as $t$ and $x_{p}$ as $x$ below. 
The initial condition for the case (I) is the configuration that 
two D-p-branes have been parallel to each other with a displacement  
$y_{0}$ until $t=0$ but are put
intersected at one angle $\theta$ at the instant $t=0$.
The setting is depicted in Fig.1.
\begin{figure}[h] 
\setlength{\unitlength}{1mm}
\begin{picture}(160,40)
\put(4,17.5){\vector(1,0){56}}
\put(30.5,8){\vector(0,1){19}}
\put(14,17.5){\vector(0,1){5.5}}
\put(14,17.5){\vector(0,-1){5.5}}
\put(4,8){\line(1,0){56}}
\put(4,27){\line(1,0){56}}
\put(4,8){\line(0,1){19}}
\put(60,8){\line(0,1){19}}
\put(4,12){\thicklines\vector(1,0){56}}
\put(4,23){\thicklines\vector(1,0){56}}
\put(29,27){\makebox(5,5){$x_{p+1}$}}
\put(54,17){\makebox(5,5){$x$}}
\put(44,28){\makebox(5,5){$Dp$}}
\put(44,2){\makebox(5,5){$Dp$}}
\put(16,17){\makebox(5,5){$y_{0}$}}
\put(31,2){\makebox(5,5){$L_{p}$}}
\put(63,15){\makebox(5,5){$L_{p+1}$}}
\put(26,33){\makebox(5,5){(i) $t<0$}}
\put(71,16.8){\thicklines\vector(1,0){5}}
\put(71,18.2){\thicklines\vector(1,0){5}}
\put(80,17.5){\vector(1,0){56}}
\put(108,8){\vector(0,1){19}}
\put(80,8){\thicklines\vector(3,1){56}}
\put(80,27){\thicklines\vector(3,-1){56}}
\put(80,8){\line(1,0){56}}
\put(80,27){\line(1,0){56}}
\put(80,8){\line(0,1){19}}
\put(136,8){\line(0,1){19}}
\put(125,18){\makebox(5,5){$\theta /2$}}
\put(125,12){\makebox(5,5){$\theta /2$}}
\put(105,27){\makebox(5,5){$x_{p+1}$}}
\put(131,17){\makebox(5,5){$x$}}
\put(126,28){\makebox(5,5){$Dp$}}
\put(126,2){\makebox(5,5){$Dp$}}
\put(107,2){\makebox(5,5){$L_{p}$}}
\put(139,15){\makebox(5,5){$L_{p+1}$}}
\put(102,33){\makebox(5,5){(ii) $t= 0$}}
\end{picture}
\caption{The initial configuration for the case (I)}
\end{figure}
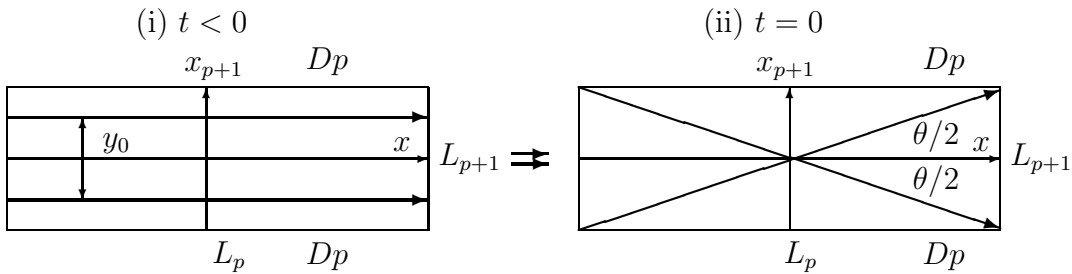

\noindent
The initial condition for the case (II) 
is the one that one of parallel two D-p-branes
were rotated by a small angle $\theta$ and are approaching each other 
with a small relative velocity $v$.
The simplest setting for each case is represented in Fig.2.
We note that $x$ is parallel to neither of the branes,
but can parametrize a world-space
of either branes, though the energy density
is $\sqrt{1+\beta^2}$ times larger then the parallel embedding
(where $\beta\equiv \tan(\theta/2)$).
Although there are multiple intersecting points in 
cases of compact dimensions, in this paper we assume that the
recombination happens only at the point $x=0$ for simplicity.
\begin{figure}[h] 
\setlength{\unitlength}{1mm}
\begin{picture}(160,40)
\put(4,17.5){\vector(1,0){56}}
\put(32,8){\vector(0,1){19}}
\put(4,8){\thicklines\vector(3,1){56}}
\put(4,27){\thicklines\vector(3,-1){56}}
\put(4,8){\line(1,0){56}}
\put(4,27){\line(1,0){56}}
\put(4,8){\line(0,1){19}}
\put(60,8){\line(0,1){19}}
\put(49,18){\makebox(5,5){$\theta /2$}}
\put(49,12){\makebox(5,5){$\theta /2$}}
\put(29,27){\makebox(5,5){$x_{p+1}$}}
\put(55,17){\makebox(5,5){$x$}}
\put(50,28){\makebox(5,5){$Dp$}}
\put(50,2){\makebox(5,5){$Dp$}}
\put(31,2){\makebox(5,5){$L_{p}$}}
\put(63,15){\makebox(5,5){$L_{p+1}$}}
\put(26,33){\makebox(5,5){$x_{p}x_{p+1}$-plane}}
\put(80,17.5){\vector(1,0){56}}
\put(106.5,8){\vector(0,1){19}}
\put(88,17.5){\vector(0,1){5.5}}
\put(88,17.5){\vector(0,-1){5.5}}
\put(80,8){\line(1,0){56}}
\put(80,27){\line(1,0){56}}
\put(80,8){\line(0,1){19}}
\put(136,8){\line(0,1){19}}
\put(115,22){\vector(0,-1){3}}
\put(115,13){\vector(0,1){3}}
\put(80,12){\thicklines\vector(1,0){56}}
\put(80,23){\thicklines\vector(1,0){56}}
\put(120,18){\makebox(5,5){$v/2$}}
\put(120,12){\makebox(5,5){$v/2$}}
\put(105,27){\makebox(5,5){$x_{9}$}}
\put(130,17){\makebox(5,5){$x$}}
\put(120,28){\makebox(5,5){$Dp$}}
\put(120,2){\makebox(5,5){$Dp$}}
\put(92,17){\makebox(8,5){$z(t=0)$}}
\put(107,2){\makebox(5,5){$L_{p}$}}
\put(139,15){\makebox(5,5){$L_{p+1}$}}
\put(102,33){\makebox(5,5){$x_{p}x_{9}$-plane}}
\end{picture}
\caption{The initial configuration for case (II) (t=0)}
\end{figure}
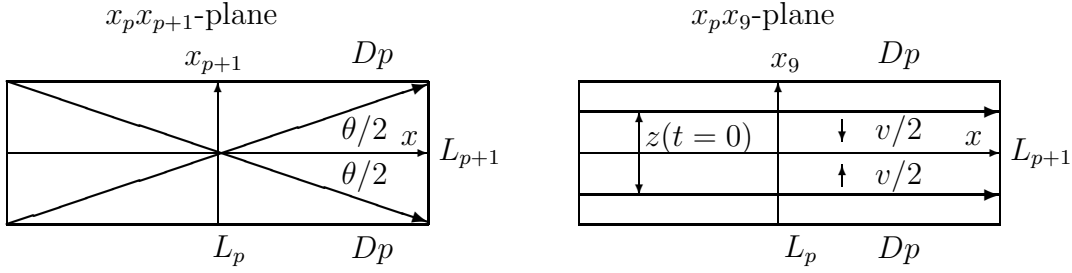

Let us give a brief review of the process.
The configuration represented as (ii) in Fig.1 cannot be distinguished from 
that of the two branes in which both are 
bent and touched at $x_{p}=x_{p+1}=0$
since they have the same energy density and flux.
The latter is unstable and each brane begins to take a shorter cut
to save the energy. This is the recombination process.
Their shapes just after that are expected to be like a hyperbola.
This instability leads to the existence of a complex tachyon field 
with a mass-squared 
\beqa
m^{2}=-\theta,\label{mass1}
\eeqa 
appearing as modes of a string connecting the two D-branes,
as shown in ref.\cite{angle}.
We will investigate the behavior of the system using 
T-dualized versions of U(2) SYM as in ref.\cite{hw}\cite{hn}. 

For each $p$, the SYM action with $x_{p+1},\cdots,x_{9}$ T-dualized 
takes the form
\beqa
S_{2Dp}=-T_{p} \int d^{p+1} x \ 
{\rm Tr} \{\ (1+)\ \frac{1}{4}F_{\mu\nu}F^{\mu\nu} +\frac{1}{2 
}(D_{\mu}X_{i})^2
- [X_{i},X_{j}]^{2}\}
\eeqa
where $T_{p}=1/g_{s}$ is the tension of a D-p-brane. 
$F_{\mu\nu}$ is the field
strength of the world-volume U(2) gauge field $A_{\mu}$ for
$\mu=0,1,..,p$. $X_{i}$ for $i=p+1,..,9$ are 
U(2) adjoint scalar fields corresponding to 
coordinates $x_{i}$ transverse to the branes. 
We note that to apply SYM, the relation 
$L_{p+1}\simeq L_{p}\theta/2 < l_{s}(=1)$ should be satisfied, 
leading to the constraint
\beqa
L_{p}<\frac{2}{\theta}.\label{lcond1}
\eeqa

We discuss dynamical behavior of the system for $t\ge 0$.
The ``background'' D-p-branes for each case for $t\ge 0$ are
represented by the configuration (VEV's):
\beqa
\begin{array}{ccc}
X_{p+1}= \left(
\begin{array}{cc}
\beta x & 0 \\
0 & -\beta x
\end{array}
\right),
 &
X_{9}= \left(
\begin{array}{cc}
z/2 & 0 \\
0 & -z/2
\end{array}
\right) ,
& X_{p+2}=\cdots =X_{8}=A_{\mu}=0.
\end{array}
\eeqa
where $\beta\equiv \tan(\theta/2)$. 
For the case (I) ($z(t)=0$),
this is T-dual 
to the configuration of
two D-(p+1)-brane with a constant field strength
$F_{p,p+1} = \beta$, as discussed in ref.\cite{hw}. 
Actually, in ref.\cite{baal}, 
tachyonic modes around the T-dualized ($F_{p,p+1} = \beta$)
background were shown to appear in off-diagonal ones of
$A_{p}$ and $A_{p+1}$ in a certain gauge,
which is T-dual to $X_{p+1}$ in the present case.
However, the connection via T-duality does not of course means
that the two systems are physically equivalent. 
(Actually, as we will show later,  
T-duality has a non-trivial effect for the system to change
of the width of a wave function by a  $\sqrt{2}$.)
Thus, we examine the spectrum from the beginning.
For case (II) ($z(t)\ne 0$),
$z/2$ is a VEV of an adjoint Higgs field
which decreases slowly.
U(2) gauge symmetry is spontaneously broken to
U(1) $\times$ U(1) by the non-trivial VEV's. 
Since we set $g_{s}$ and $\theta$
very small, the force between the two branes is so
weak\cite{jabbari}\cite{pol}(and for short distance, see
\cite{oneangle})
that we can regard the velocity as a constant, leading to
\beqa
z=z_{0} -v t.
\eeqa
In this case  
potentially tachyonic modes should appear 
in the off-diagonal elements of $A_{p}$ and $X_{p+1}$ again.
In addition, there might be  a possibility
that tachyonic modes appear in off-diagonal elements of 
$A_{0}$ and $X_{9}$ because the two D-p-branes
approaching with a small velocity is 
T-dual to a two D-(p+1)-brane system
with a constant field strength (though it is electric),
and is also 
considered to be a Wick-rotated version of 
branes at one angle $v$\cite{pol}.
So, we denote the fluctuations, 
including those of $A_{0}$ and $X_{9}$, as
\begin{equation}
\begin{array}{ccc}
A_{\mu}= \left(
\begin{array}{cc}
0 & c_{\mu}^{*} \\
c_{\mu} & 0
\end{array}
\right), &
X_{p+1}= \left(
\begin{array}{cc}
0 & d^{*} \\
d & 0 
\end{array}
\right) 
, &
X_{9}= \left(
\begin{array}{cc}
0 & w^{*} \\
w & 0 
\end{array}
\right)
\end{array}\label{ansatz2}
\end{equation}
(and denote in particular, $c_{p}$ as $c$), and
discuss their spectra below.

The part of the action second order in the fluctuations is
\beqa
S_{2Dp}|_{{\rm 2nd}}
&=&T_{p} \int d^{p+1}x [ 
-\partial_{\mu}c^{*}_{\nu}\partial^{\mu}c^{\nu} 
+|\partial_{\mu}c^{\mu}|^2 
-\partial_{\mu}d^{*}\partial^{\mu}d 
-\partial_{\mu}w^{*}\partial^{\mu}w\nonumber\\ 
& &+2 i \beta (c^{*}d -c d^{*})
-2 i \beta x(c^{*}_{\mu}\partial^{\mu}d 
-c_{\mu} \partial^{\mu}d^{*}) 
+iv(c_{0}^{*}w-c_{0}w^{*})\nonumber\\
& &-iz(c^{*}_{\mu}\partial^{\mu}w 
-c_{\mu} \partial^{\mu}w^{*})
-(4(\beta x)^{2}+z^{2})|c_{\mu}|^{2}-|2\beta x w-zd|^{2}
].\label{secondaction22}
\eeqa 
We consider such a gauge fixing condition 
that it is written as $D_{\mu}^{(0)}A^{\mu}\equiv\partial_{\mu}
A^{\mu}-i[A_{\mu}^{(0)},A^{\mu}]=0$ in a T-dualized pure 
Yang-Mills system, which gives
\beqa
\partial_{\mu}c^{\mu}=i(2\beta x d+zw).\label{gf3}
\eeqa
This is a generalized version of the condition
chosen in ref.\cite{baal}\cite{hw}.
Integrating by part some terms and  using (\ref{gf3}),
we have 
\beqa
S_{2Dp}|_{{\rm 2nd}}&=&  T_{p} \int d^{p+1}x [
-c^{*}_{\nu} \{ -(\partial_{\mu})^{2}
+4(\beta x)^{2} +z^{2}\} c_{\nu} 
- d^{*}  \{ -(\partial_{\mu})^{2}
+4(\beta x)^{2} +z^{2} \}d \nonumber\\
& &-w^{*}\{ -(\partial_{\mu})^{2}
+4(\beta x)^{2}+z^{2} \} w
-4 i \beta (c^{*}d -c d^{*})-2iv(c_{0}^{*}w-c_{0}w^{*})
].\label{secondaction23}
\eeqa
With the redefinition of the fields as
\beqa
\tilde{c} \equiv \frac{c+id}{\sqrt{2}} ,\ 
\tilde{d} \equiv \frac{d+ic}{\sqrt{2}}, 
\label{defctilder} 
\eeqa
the term $-4 i \beta (c^{*}d -c d^{*})$ in (\ref{secondaction23})
give rise to their mass terms
$-(-4\beta)|\tilde{c}|^{2}-(4\beta)|\tilde{d}|^{2}$.

Here we set $z(t)=v=0$ and discuss the case (I) for a while. 
Let us consider mode expansions of the fields $\tilde{c}$, 
$\tilde{d}$ and $c_{\hat{\mu}}$ ($\hat{\mu}=0,1,\cdots,p-1 $)  
with respect to $x$ and $x_{\bar{\mu}}$ 
($\bar{\mu}=1,\cdots,p-1$) (the latter for D-p-branes with $p\ge 2$) 
to examine their spectra.
As for $x_{\bar{\mu}}$, we use $u_{k}(x_{\bar{\mu}})\equiv 
e^{ik_{\bar{\mu}}x^{\bar{\mu}}}$,
free wave functions with a wave vector $ k_{\bar{\mu}}$.
As for $x$, the operator for each field is written as
\beqa
[-(\partial_{x})^{2}+4(\beta x)^{2} -4 a \beta]\label{ho}
\eeqa 
where $a=1,-1,0$ for $\tilde{c}$, $\tilde{d}$ and 
$c_{\hat{\nu}}$, respectively.
If it were not for the periodicity of $x$,
the eigenvalues and eigenfunctions could be given by
$\tilde{m}^{2}_{n}$ and $f_{n}$ as
\beqa 
\tilde{m}^{2}_{n}&=&2\beta(2 n -a)\label{massform}\label{eigenv}\\
f_{n}&=&\frac{1}{\sqrt{2^{n}n!}}(\frac{2\beta}{\pi})^{1/4}
e^{-\beta x^{2}}H_{n}(\sqrt{2\beta}x) \label{eigenset}
\eeqa
for each non-negative integer $n$ where $H_{n}$ are Hermite polynomials;
since $x$ direction is compactified,   
exact eigenfunctions are expected to be mathematically complicated, 
e.g. related to theta functions
as in ref.\cite{baal}\cite{hw}.
However, we can see that $f_{n}$ in (\ref{eigenset}) is localized around 
$x=0$ with a width 
$\delta\equiv 1/\sqrt{2\beta}\simeq 1/\sqrt{\theta}$.
So, if one consider 
$L_{p}$ such that
\beqa
\frac{1}{\sqrt{\theta}} \ll L_{p},\label{lcond2}
\eeqa 
one can use (\ref{eigenv}) and (\ref{eigenset})  
as an approximate set of eigenvalues and eigenfunctions.
Then, we can see that
the lowest mode of $\tilde{c}$ (we denote as $\tilde{c}_{0}$)
is a complex tachyon field with a mass-squared
\beqa
(m_{0})^{2}=-2 \beta,
\eeqa 
which agrees with (\ref{mass1}) obtained in ref.\cite{angle} 
for a small $\theta$, while
the lowest modes of $c_{\bar{\nu}}$ are massless and the others are
massive. 
We  note that the condition (\ref{lcond2}) is compatible with
the one (\ref{lcond1}) if $\theta \ll 1$. 
So, we consider such a range of  $L_{p}$ throughout this paper.
(We set $\theta\simeq 2\beta=10^{-2}$ and $L_{p}=100$ later.)
We also note that the approximate width of $f_{n}$
is larger than the eigenfunction in the T-dualized theory
in ref.\cite{hw}\cite{baal} by  a factor $\sqrt{2}$, 
which may be one of non-trivial effects of the T-duality. 

Expanding $\tilde{c}$ and the other fields such as 
\beqa
\tilde{c}(x_{\mu}) =\sum_{n}\int\frac{d^{p-1}{\bf k}}{(2\pi)^{p-1}} 
\tilde{c}_{n,k_{\bar{\mu}}} (t) 
u_{k}(x_{\bar{\mu}})f_{n}(x), 
\label{expansion2}
\eeqa
we obtain the action 
\beqa
S_{2Dp}|_{{\rm 2nd}}=T_{p}\int d t
\sum_{n,k_{\hat{\mu}}} \{ |\frac{d\tilde{c}_{n,k}}{dt}|^{2}
-\omega_{n,k}^{\ \ \ 2}|\tilde{c}_{n,k}|^{2}
+(\tilde{c}_{n,k}\to \tilde{d}_{n, k}, \ c_{\hat{\nu},n,k})\}
\label{secondaction3}.
\eeqa
where $\omega_{n,k}^{\ \ \ 2}$ for $t\ge 0$ is given as
\beqa
\omega_{n,k}^{\ \ \ 2}= k^{2}+2 \beta (2n-a)\label{omega1}
\eeqa
where $k\equiv \sqrt{(k_{\bar{\mu}})^{2}}$.
(Note that $a=1,-1,0$ for $\tilde{c}_{n,k}, \tilde{d}_{n,k},  
c_{\hat{\nu},n,k}$, 
respectively.)
This looks much like the action of a collection of harmonic 
oscillators
with a mass $T_{p}$, although it includes modes with negative 
frequency-squareds; we can easily see that
the modes $\tilde{c}_{0,k}$ with 
$0 \le k^{2} \le 2\beta$
have negative $\omega^{2}$'s (while all the other modes 
have non-negative $\omega^{2}$'s). 
Thus, they are the tachyonic modes in the case (I) with $p\ge 2$.
For the case of D-strings, there is no d.o.f. of $k_{\bar{\mu}}$ and
$\tilde{c}_{0}$ is the only (complex) tachyonic mode with  
$(\omega_{0})^{2}=-2\beta $ for $t\ge 0$.

For $t<0$, one should set
$2\beta =\theta =0$ and instead turn on a displacement of the branes
$y_{0}$ in $x_{p+1}$ direction. $y_{0}$ contribute to the
second order action as a mass-squared for each of the off-diagonal
fluctuations. Since the direction of $x_{p}$ is compactified,
the momentum component $k_{p}$ is discretized as 
$k_{p}=2\pi n_{p}/L_{p}$ 
for an integer $n_{p}$, and the mode functions is
\beqa
g_{n_{p}}(x)=e^{ik_{p}x}/\sqrt{L_{p}}.
\eeqa
The modes  
$\tilde{c}_{n_{p},k}^{(0)}\equiv
(c_{n_{p},k}^{(0)}+id_{n_{p},k}^{(0)})/\sqrt{2}$,
corresponding to $g_{n_{p}}(x)$,
have such $(\omega_{n_{p},k}^{(0)})^{2}$  as
\beqa
\omega_{n,k}^{(0)\ \ 2}=(y_{0})^{2}+(k_{\bar{\mu}})^{2}
+(2\pi n_{p}/L_{p})^{2}\label{omega01}
\eeqa
(where we attach the index $(0)$ to the qualities defined for $t<0$).

Let us move on to the case (II) where $z\ne 0$.
As seen in (\ref{secondaction23}), $z^{2}$ merely plays a role of 
a mass-squared for each field.
Making mode-expansions of $\tilde{c},\tilde{d}$, and
$c_{\hat{\mu}}$ in terms of $u_{k}$ and $f_{n}$,
we have the action (\ref{secondaction3}) with frequencies
\beqa
\omega_{n,k}^{'\ \ \ 2}= k^{2}+2 \beta (2n-a)+(z_{0}-v t)^{2}
\label{omega2}
\eeqa
where $a=1,-1,0$ for $\tilde{c}_{n,k} \tilde{d}_{n,k},
c_{\bar{\nu},n,k}$, respectively again.
As we will discuss later,
we need the fact that all the modes have positive frequency-squareds 
at $t=0$, so we set $z_{0}=0.5 l_{s}$, relatively a large value
but not to exceed the applicable range of SYM.
Since $z(t)$ decreases gradually to zero,
again (potentially) tachyonic modes are 
the modes $\tilde{c}_{0,k}$ with $k^{2}\le 2\beta$ for the case with
$p\ge 2$, and $\tilde{c}_{0}$ is the only (complex) tachyonic mode
for the case of D-strings.

Qualitative behavior of the system in the case (II) is as follows:
at $t=0$, all the modes have positive $\omega^{2}$'s, hence have
some small fluctuations.
As $z(t)$ decreases to $z=\sqrt{2\beta}$ (which gives
$\omega_{0,0}^{'\ \ 2}=0$), $\tilde{c}_{0,0}$ starts to blow up, 
and
so do the modes $\tilde{c}_{0,k}$'s with a small $|k|$ one after
another when   
$z$ becomes a critical displacement $z^{*}_{k}$ 
which gives $\omega_{0,k}^{'\ \ 2}=0$.
The blow-ups together give rise to the tachyon condensation
(to cause recombination).

In the next section we will focus on
the potentially tachyonic modes $\tilde{c}_{0,k}$ with $2\beta \gg v$,
and make a quantum-mechanical analysis
using the action (\ref{secondaction3}) with the obtained $\omega^{2}$'s.

We note that one can regard $c_{0}$ and $w$ as non-tachyonic fields for
the parameter $2\beta \gg v$.
As for the fields $c_{0}$ and $w$,
the last term in (\ref{secondaction23})
might also seem to produce mass terms of linear combination of 
$c_{0}$ and $w$, 
in the same way as the case of $c$ and $d$,
with $2\beta$, $c$,$d$ replaced by $v$,$c_{0}$, $w$, respectively, 
resulting in a complex tachyonic field.
However, the sign of the kinetic term of $c_{0}$
is different from that of $w$, and, if one carry out 
the diagonalization, it results in a non-local action including 
square roots of differential operators;
it is unclear whether $c_{0}$ and $w$ include tachyonic modes.
Let us consider here the case with $2\beta \gg v$.
Then, even if there would be tachyon modes composed of
$c_{0}$ and $w$, 
the critical distances $z'_{*}$'s of them are much shorter than
those of $\tilde{c}$, and
it is much before the postulated tachyons of 
$c_{0}$ and $w$ get ``awake"
that the tachyon modes of $\tilde{c}$ condensate and
the recombination proceeds.
Thus, we consider the parameter region $2\beta \gg v$ 
and deal with $c_{0}$ and $w$ as non-tachyonic fields.  


\section{A quantum-mechanical analysis on tachyon condensation}
\setcounter{equation}{0}
\setcounter{footnote}{0}

In this section we analyze in detail how the tachyon fields condense,
applying quantum mechanics to the action (\ref{secondaction3}). 
We assume that the system is initially 
at zero temperature, i.e. at the ground state, for simplicity,
though the essential point of our argument will not be changed  
as long as the temperature is low.

One of the two basic ingredients (we will use here) is
that in a harmonic oscillator with a positive constant $\omega^{2}$, 
the ground state wave function is Gaussian, and 
its variable $q$ has a typical amplitude (absolute value) of fluctuation
$q_{{\rm typical}}\equiv \sqrt{<q^{2}>}=1/\sqrt{2m
\omega}=\Delta/\sqrt{2}$ 
where $<A>$ denotes a VEV of an operator $A$,
$m$ is a mass of $q$, and $\Delta $ is a width of $q$'s wave function.
Another basic one is that even if $\omega^{2}$ is time-dependent and 
becomes negative afterward, time-evolution of its wave function 
and hence that of $q_{{\rm typical}}$ can be deduced within WKB 
(second order) approximation,
by using its propagator or kernel, as in e.g. ref.\cite{fey}.
Taking these two into account, time-evolution behavior of the 
typical amplitude of each mode can be evaluated using quantum mechanics,
if an initial condition is such that 
$\omega^{2}$ of the mode is positive at $t=0$ and 
change into a negative one later:
the initial typical value is $\sqrt{g_{s}/2\omega(t=0)}$ since
in this case the tension $T_{p}=1/g_{s}$ corresponds to the mass.
Since the wave function at $t=0$ is Gaussian,
the absolute value of the wave function at any $t$ 
remains to be Gaussian 
within WKB (second order) approximation, and 
the system is characterized only by its width $\Delta(t)$,
which is equal to 
a typical value of the variable $q(t)_{{\rm typical}}$
times $\sqrt{2}$.
(These are reviewed in the appendix A. See ref.\cite{fey} 
for more detail.)
We note that typical amplitudes of the modes with non-negative 
$\omega^{2}$'s are 
suppressed by $\sqrt{g_{s}}$ 
in their numerators as 
$|\tilde{c}_{n, k}(t)|_{{\rm typical}} =\sqrt{g_{s}/2 \omega(t) }
\ \ (n\ge 1)$.
We also note that the WKB approximation is quite good 
until tachyonic modes blow up because
higher order terms are also suppressed by an extra factor of $O(g_{s})$.

If we define a typical amplitude of fields as 
$|\tilde{c}(x_{\mu})|_{{\rm typical}}\equiv \sqrt{<|\tilde{c}(x_{\mu})|^{2}>}$,
we have to sum
$|\tilde{c}_{n, k}|_{{\rm typical}}^{2} (f_{n})^{2}|u_{n,k}|^{2}$ 
with respect to $n$ and integrate it with respect to 
$k_{\bar{\mu}}$ except for the case of D-strings,
but the integral is divergent.
However, this is a kind of divergence that appears 
in a usual quantum field theory
and has nothing to do with the blow-up of tachyons.
Thus, we ignore the contribution of all the non-tachyonic modes 
and define the typical amplitude of
$\tilde{c}(x_{\mu})$ in the case of D-p-branes with $p\ge 2$ as
\beqa
|\tilde{c}(x_{\mu})|_{{\rm typical}} \equiv \{ \int_{k^{2}\le 2\beta} 
\frac{d^{p-1}{\bf k}}{(2\pi)^{p-1}}<|\tilde{c}_{0, k}(t)|^{2}> 
(f_{0})^{2}\}^{1/2}\equiv A(t)f_{0}\label{typfldef}
\eeqa
where $A(t)$ is a time-dependent part of $|\tilde{c}(x)|$ and $f_{0}(x)=
(2\beta/\pi)^{1/4}e^{-\beta x^{2}}$.
In the case of D-strings 
we define 
\beqa
|\tilde{c}(x_{\mu})|_{{\rm typical}}
\equiv\sqrt{<|\tilde{c}_{0}(t)|^{2}>}\equiv  A(t)f_{0}.
\label{typfldef2}
\eeqa
In this paper we regard the quantity (\ref{typfldef}) 
or (\ref{typfldef2}) as a VEV of $|\tilde{c}(x)|$ due 
to tachyon condensation.
As in ref.\cite{fey} and the appendix A, 
the time-dependent width $\Delta_{k}(t)$ of the wave function of 
$\tilde{c}_{0, k}(t)$, which is equal to 
$\sqrt{2}|\tilde{c}_{0,k}(t)|_{{\rm typical}}$,
can be written in terms of 
the initial width $\Delta_{k}(0)=\sqrt{g_{s}/\omega_{k}(t=0)}$ 
and two independent solutions 
to the equations of motion 
\beqa
\frac{d^{2} \tilde{c}_{0,k}}{dt^{2}}= 
-\omega_{0,k}^{2} \tilde{c}_{0,k}.
\label{eom1}
\eeqa
We denote the solutions as ``basis functions'' .
Below we solve the equation of motion for each case
and use the basis functions and $\Delta_{k}(0)$
to evaluate $|\tilde{c}(x_{\mu})|_{{\rm typical}}$.

For the case (I) with $p\ge 2$,
we can easily see that 
the basis functions for $t\ge 0$ are
$\cosh (Kt)$ and $\sinh (Kt)$
where $K\equiv \sqrt{-\omega^{2}_{0,k}}=\sqrt{2\beta-k^{2}}$ 
($\ge 0$ for tachyonic modes),
and the initial width for $\tilde{c}_{n_{p},k}^{(0)}$ ($t\le 0$)
is $\Delta_{n_{p},k}^{(0)}(0)
=\sqrt{2}|\tilde{c}_{n_{p},k}^{(0)}(t=0)|_{{\rm typical}}=
\sqrt{g_{s}/\omega_{n_{p},k}^{(0)}}$
where $\omega_{n_{p},k}^{(0)}$ is given in (\ref{omega01}). 
However, the mode (eigen-) functions (of the operator 
with respect to $x$ arising in the action) 
change at $t=0$ from $g_{n_{p}}(x)$ to $f_{n}(x)$, so we have to ``translate'' 
$|\tilde{c}_{0,k}^{(0)}(t=0)|_{{\rm typical}}$ 
into $|\tilde{c}_{0,k}(t=0)|_{{\rm typical}}$.
(We note that $\tilde{c}_{0,k}$ is related only to  
$\tilde{c}_{n_{p},k}^{(0)}$ with the 
same $k_{\bar{\mu}}$
since the system keeps translation symmetry in $x_{\bar{\mu}}$
directions
even at $t=0$.)
If we define $N_{n,n_{p}}\equiv \int dx f_{n}^{*} g_{n_{p}}$,
it holds $g_{n_{p}}=\sum_{n}N_{n,n_{p}}f_{n}$ and we have 
$\tilde{c}_{n,k}=\sum_{n_{p}}N_{n,n_{p}}c_{n_{p},k}^{(0)}$,
and hence 
\beqa
\Delta_{k}(t=0)^{2}=2|\tilde{c}_{0,k}(t=0)|^{2}_{{\rm typical}}
=\sum_{n_{p}}|N_{0,n_{p}}|^{2}\frac{g_{s}}{\omega_{n_{p},k}^{(0)}}.
\label{delta01}
\eeqa
With the $\Delta_{k}(t=0)^{2}$, the typical value 
is given through (\ref{formaldelta}) as
\beqa
|\tilde{c}_{0,k}|_{{\rm typical}}^{2}=\frac{\Delta_{k}(t)^{2}}{2}=
\frac{\Delta_{k}(0)^{2}}{2}\cosh^{2}Kt
+\frac{g_{s}^{2}}{2\Delta_{k}(0)^{2}K^{2}}\sinh^{2}Kt, 
\label{deltak1}
\eeqa
and we can obtain the typical amplitude $|\tilde{c}(x_{\mu})|$ as
\beqa
|\tilde{c}(x_{\mu})|_{{\rm typical}}^{2}=(A(t)f_{0})^{2}=
\frac{\Omega_{p-2}}{(2\pi)^{p-1}}
\int_{0}^{\sqrt{2\beta}} k^{p-2}dk 
\frac{\Delta_{k}(t)^{2}}{2}(f_{0}(x))^{2}
\label{staticcfinal1}
\eeqa
where $\Omega_{p-2}$ is the volume of (p-2)-sphere.
This is the formula for time-evolution of the typical amplitude of
the tachyonic fluctuation for the case (I) with $p\ge 2$.
For $p=1$ (D-strings) case, we have
\beqa
|\tilde{c}_{0}(x_{\mu})|_{{\rm typical}}^{2}\equiv (A(t)f_{0})^{2} 
=\frac{\Delta_{k=0}(t)^{2}}{2}(f_{0})^{2}\label{staticstring1}
\eeqa
where $\Delta_{k}(t)$ is given in (\ref{deltak1}).

For all value of $p$,
order of $|\tilde{c}(x^{\mu})|_{{\rm typical}}$ in terms of $g_{s}$
is ${\cal O}(g_{s})$.  In addition, $|\tilde{c}(x^{\mu})|_{{\rm typical}}$
has x-dependence through the Gaussian function $f_{0}(x)$.
This means that the excitation of the tachyon is localized around $x=0$
as in ref.\cite{baal}\cite{hw}\cite{hn}
though the width is relatively broad such as $\delta=1/\sqrt{\beta}$.

A rough behavior of $A(t)$ can be read from the factor
$e^{\sqrt{2\beta-k^{2}}t} \sim  e^{\sqrt{2\beta}t}$ in
(\ref{staticcfinal1}) or (\ref{deltak1}).
This can be inferred from the classical solution to the equation of
motion, and a typical time scale is defined as $T^{(I)}\equiv
1/\sqrt{2\beta}$.
The formula for $A(t)$ is a bit complicated, but 
if we set $L_{p}=100l_{s}$, we have
$\sum_{n_{p}=-2}^{2}|N_{n,n_{p}}|^{2} \simeq 0.9987$, so
it suffices to have the terms with $n_{p}=0,\pm 1, \pm 2$ in
(\ref{delta01}).
If we set $y_{0}=0.5 l_{s}$, relatively a large value,
we can approximate $\omega_{n_{p},k}\simeq y_{0}$ for the $n_{p}$'s, 
and hence $\Delta_{k}(t=0)^{2}\simeq g_{s}/y_{0}$.
So, we have
\beqa
|\tilde{c}_{0,k}|^{2}_{{\rm typical}}\simeq
\frac{g_{s}y_{0}}{2}\{
\frac{1}{2\beta-k^{2}}\sinh^{2}\sqrt{2\beta-k^{2}}t
+\frac{1}{(y_{0})^{2}}\cosh^{2}\sqrt{2\beta-k^{2}}t
\}.\label{staticcfinal}
\eeqa
Since $y_{0}\gg 2\beta$ now, the first term in
(\ref{staticcfinal}) is dominant.
The time-dependence of $A(t)$'s for $p=1,2,4$ is plotted in Fig.3, for
a specific value
of the angle $\theta\simeq 2 \beta =0.01$ and $g_{s}=0.01$, which we set
on the basis of the brane phenomenology
(e.g. refs.\cite{oneangle}\cite{tye1}).
\begin{figure}[h]
\begin{center}
\includegraphics[width=8cm]{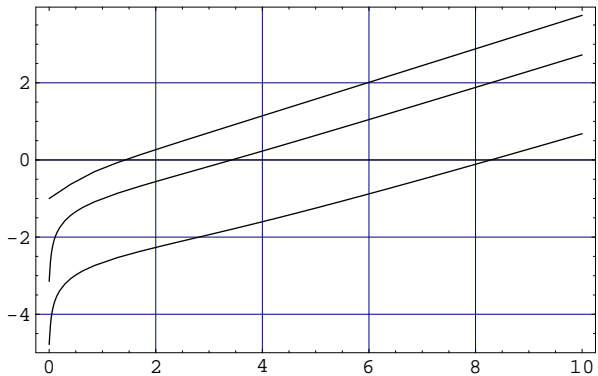}
\caption{
%
$\log_{10}A(t)$ vs. $t/T^{(I)}$.The topmost
graph is 
for D-strings, the middle is for D-2-branes and the lowest is for
D-4-branes, with 
$\theta\simeq 2\beta=0.01$, $g_{s}=0.01$ and $T^{(I)}=10 l_{s}$.
}
\end{center}
\end{figure}
We can read a tendency from (\ref{staticcfinal1}) or Fig.3
that the blow-up is more
delayed for the case with a larger value of $p$.
This fact can be explained as follows: 
the tachyonic modes with less momentum tend to blow up more rapidly,
but there are larger number of tachyonic modes 
for larger momentum.
That is, except for the case of D-strings,
for larger value of $p$,
the most ``active'' tachyonic mode which has $k=0$ is 
more smeared in some region of $k$  
and the start of the blow up becomes slower.

Next we derive $A(t)$ for the case (II). 
The frequency-squared for $\tilde{c}_{0,k} $ is
\beqa
\omega_{0,k}(t)^{2}= (z_{0}-v t)^{2} +k_{\tilde{\mu}}^2-2 \beta.
\label{omega3}
\eeqa
We note that each $\omega_{0,k}^{\ \ 2}$ is positive at $t=0$ for 
$z_{0}=0.5 l_{s}$.
The behavior of the system in terms of its wave function is 
as follows: 
As $z(t)$ decreases gradually,
so does each $\omega_{0,k}^{\ \ 2}$ and
the wave function $\Psi(t,\tilde{c}_{0,k})$ of 
$\tilde{c}_{0,k}$ spreads little by little.
At $t= t_{0}\equiv (z_{0}-\sqrt{2\beta})/v$, 
$\omega^{\ \ 2}_{0,0}$ 
(with the lowest momentum $k=0$) becomes zero and 
then negative, i.e.  tachyonic, and the corresponding
wave function begins to spread radically, resulting in
the blow-up of the mode.
After that, the modes with lower $k$ are also getting tachyonic  
and their wave functions also start to spread one after another, 
leading to the 
tachyon condensation.  

To obtain $A(t)$ in this case, we have only to find two solutions to
(\ref{eom1}) with (\ref{omega3}) and substitute it for (\ref{formaldelta}).
However, its exact solutions is a bit difficult to deal with,
so we find some approximate solutions:
Let us define
$t_{*}(k)\equiv (z_{0}-\sqrt{2\beta-k^{2} })/v$
which gives
$\omega_{0,k}(t_{*})=0$,
and rewrite the $\omega_{0,k}^{2}$ using $t_{*}$ as
\beqa
\omega_{0,k}(t)^{2}= -\alpha_{k} (t-t_{*})+v^{2}(t-t_{*})^{2}
\label{omega22}
\eeqa
where we denote the coefficient $\alpha_{k} \equiv \sqrt{8 v^{2} \beta'}$
and $\beta'\equiv \beta-k^{2}/2$.
If $t$ approaches $t_{*}$, the second term quadratic in $t-t_{*}$ of
(\ref{omega22}) becomes so small as to be neglected compared to
the first term. (We denote the time as $t=t_{1}$.) 
Thus, for $t>t_{1}$, two Airy functions can be used as basis functions.
On the other hand, taking a closer look will let one notice that 
(\ref{eom1}) with (\ref{omega22}) has the same form  
as the Schrodinger equation 
in one dimension with relatively slowly-changing potential
(if we replace $t$ by $x$). 
So, when $\omega_{0,k}(t)^{\ \ 2}$ is relatively large 
and changing slowly, 
we can apply ``WKB approximation''
to get two approximate basis functions. 
Thus, if this approximation can be applied until the time
$t=t_{1}$, we can follow the time-evolution of 
$\Psi(t,\tilde{c}_{0,k})$ and  $\Delta_{k}(t)$
by using the ``WKB-approximated'' solutions until $t=t_{1}$, and then
letting the Airy functions take over the role of basis functions. 
We take this prescription to evaluate their 
time-evolution.

The ``WKB-approximated'' basis functions are
\beqa
\phi_{1}(t)=\omega_{0,k}^{-1/2}\cos(\int^{t}_{t_{1}}dt'
\omega_{0,k}(t')),\nonumber\\
\phi_{2}(t)=\omega_{0,k}^{-1/2}\sin(\int^{t}_{t_{1}}dt'
\omega_{0,k}(t')).\label{basis1}
\eeqa
This approximation is valid  as far as
\beqa
v\ll |\omega_{0,k}|^{2}{\rm \ \ and \ \ } 
|\frac{\partial\omega_{0,k}}{\partial t}|\ll |\omega_{0,k}|^{2}.
\label{cond1} 
\eeqa
On the other hand, the condition
for the Airy functions to be used is
\beqa
|t-t_{*}|\ll \frac{\sqrt{8\beta'}}{v}.\label{cond2}
\eeqa
So, if we choose $t_{1}\equiv t_{*}-\sqrt{8\beta'}/10 v$ 
to satisfy (\ref{cond2}) (please note the sign),
the condition to satisfy (\ref{cond1}) is written as
\beqa
v\ll 2\beta'.\label{cond3}
\eeqa
which is compatible with the condition to regard $c_{0}$ and $w$
as only massive fields.
So, in this paper we consider such a system.
One may worry that tachyonic modes with large $k$
do not satisfy (\ref{cond3}) for any set of $\beta$ and $v$.
In fact, in a set-up with $v\ll 2\beta$,
$v$ is so slow that the tachyonic fluctuation $\tilde{c}$
blows up enough much before the tachyon modes with large $k$  
``wake up''. So, the above point 
is not practically a problem.

Using the Airy functions $Ai(\tau)$ and $Bi(\tau)$,
we obtain the width $\Delta_{k}(t)^{2}$ of 
$|\Psi(\tilde{c}_{0,k},t)|^{2}$ as
\beqa
\Delta_{k}(t)^{2}&=&g_{s}[  \frac{\pi^{2}}{\omega_{1}}
\{Bi'(\tau_{1})Ai(\tau)-Ai'(\tau_{1})Bi(\tau)\}^{2}\nonumber\\
& &+\frac{\omega_{1}\pi^{2}}{(8v^{2}\beta')^{1/3}}
\{Bi(\tau_{1})Ai(\tau)-Ai(\tau_{1})Bi(\tau)\}^{2}
],\label{movingck}
\eeqa
where $Ai'(\tau)=\frac{dAi(\tau)}{d\tau}$, 
$\tau\equiv \alpha^{1/3}_{k} (t-t_{*})$,  
$\omega_{1}\equiv
\omega_{0,k}(t=t_{1})(=\frac{\sqrt{11}}{10}\sqrt{8\beta'})$ 
and 
$\tau_{1}\equiv \alpha^{1/3}_{k} (t_{1}-t_{*})$.
And we can obtain $|\tilde{c}(x^{\mu})|_{{\rm typical}}$
for $p\ge 2$ for case (II) by substituting
(\ref{movingck}) for (\ref{staticcfinal1}).
For the case of D-strings, $|\tilde{c}(x^{\mu})|=(\Delta_{k=0}(t)f_{0})^{2}/2$
where $\Delta_{k=0}(t)$ is given in (\ref{movingck}).

As in the case (I), the blow-up of the tachyonic fluctuation
is localized around $x=0$ with the  width $\delta=1/\sqrt{\beta}$.
A rough behavior of $A(t)$ is $\sim 
\exp [ \frac{2}{3} \{(8v^{2}\beta)^{1/6}(t-t_{*})\}^{3/2} ]$, 
so we can see that a typical time scale of the decay is
$T^{(II)}\equiv 1/(8v^{2}\beta)^{1/6}\cong 1/(4v^{2}\theta)$.
That is, the typical time scale depends more strongly on the relative
velocity of the branes than on the angle $\theta$.
The reason for this is that
after reaching the critical distance, 
the larger the velocity is, the earlier potentially tachyonic modes
``wake up'' to become truly tachyonic and blow up.
Time-dependent behavior of
$A(t)$ in the specific case of D-4-branes with 
$2\beta=10^{-2}$ and $v=10^{-4}$ is plotted in Fig.4
where $T^{(II)}\simeq 36.84 l_{s}$. 
\begin{figure}[h]
\begin{center}
\includegraphics[width=8cm]{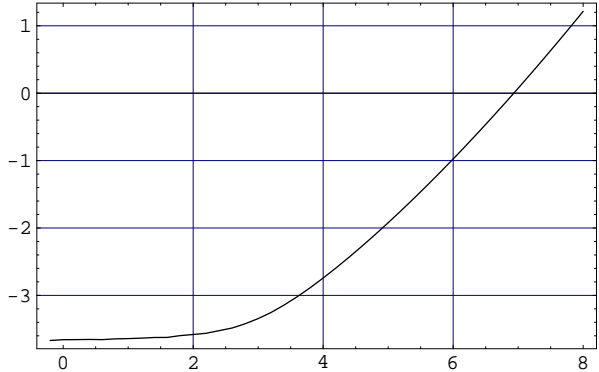}
\caption{
%
$\log_{10}A(t)$ vs. 
$(t-t_{0})/T^{(II)}$ for D-4-branes; $2\beta=10^{-2},\ v=10^{-4}$ and
 $g_{s}=0.01$.
}
\end{center}
\end{figure}

In this way, we have obtained, for both cases, explicit functions
of typical amplitudes of tachyonic fluctuations
$|\tilde{c}(x_{\mu})|_{{\rm typical}}=A(t)f_{0}$.

For convenience,
let us define the following quantity for a function of $k$, $g(k)$:
\beqa
<<g(k)>>\equiv \int \frac{d^{p-1}{\bf k}}{(2\pi)^{p-1}}
\frac{\Delta_{k}(t)^{2}}{2} g(k)\label{gk1}
\eeqa
where it is normalized as $<<1>>=A(t)^{2}$.
For a later use, we also define a ``typical amplitude of 
tachyon's wave vector'' of tachyonic fluctuations
as $\sqrt{\overline{k^{2}}}
\equiv (<<k^{2}>>/<<1>>)^{1/2}$.
By $\sqrt{\overline{k^{2}}}$ we can estimate 
which momentum-possessing tachyon modes 
contribute the most to the blow-up, 
since the weight for a tachyon mode with a momentum ${\bf k}$ 
in $|\tilde{c}(x^{\mu})|^{2}$ 
is $(\Delta_{k})^{2}/2$, as in (\ref{typfldef}) with
$|\tilde{c}_{0,k}|_{{\rm typical}}=\Delta_{k}(t)^{2}/2$.

\section{Time-evolution of the D-p-branes' shapes}
\setcounter{equation}{0}
\setcounter{footnote}{0}

In this section we discuss time-evolution  
of the D-p-branes' shapes and show seemingly queer behavior 
of  the brane's shape occurs. 
Then, we  give a physical interpretation of the behavior 
to conclude that it is not actually queer.

With (\ref{defctilder}) and (\ref{expansion2}), 
the typical {\it value} of $d(x_{\mu})$ is  
\beqa
d(x_{\mu})_{{\rm typical}} \cong 
-i\frac{\tilde{c}(x_{\mu})_{{\rm typical}} }{\sqrt{2}}=
\frac{-i}{\sqrt{2}}
\int_{k^{2}\le 2\beta}\frac{d^{p-1}{\bf k}}{(2\pi)^{p-1}}
\tilde{c}_{0,k}(t)_{{\rm typical}}
u_{k}(x_{\bar{\mu}})f_{0}(x)
\label{dbehavior}
\eeqa
since $\tilde{d}_{{\rm typical}}$ is negligible.
We note that each $\tilde{c}_{0,k}(t)_{{\rm typical}}$
has its phase as $\tilde{c}_{0,k}(t)_{{\rm typical}}=
|\tilde{c}_{0,k}(t)|_{{\rm typical}}e^{i\delta_{k}}$,
and so does $d(x_{\mu})_{{\rm typical}}$.
We regard this quality as the VEV of the field $d(x_{\mu})$.
Then, the transverse scalar U(2) field $X_{p+1}$
has the VEV of the form: 
\beqa
X_{p+1}= \left(
\begin{array}{cc}
\beta x & d^{*}_{{\rm typical}} \\
d_{{\rm typical}} & -\beta x
\end{array}
\right).\label{xp1}
\eeqa

Let us get geometrical information on the branes 
from (\ref{xp1}), first for the case (I). 
In this case there is only one transverse scaler field 
which has non-trivial VEV elements, and, all the scalar fields are 
commutative with each other. 
Thus, as done in such conventional cases and in ref.\cite{hn},
it is a logical step 
to choose the gauge to diagonalize the VEV's of 
the scalar field and interpret the diagonal parts as
the positions of the two branes.
$X_{p+1}$ is diagonalized as
\begin{equation}
\begin{array}{cc}
X^{'}_{p+1}=U_{0} X_{p+1} U_{0}^{-1} 
&=
\left(
\begin{array}{cc}
y(x,t) & 0 \\
0 & -y(x,t)
\end{array}
\right) 
\end{array}\label{diagonal1}
\end{equation}
where $y=\sqrt{(\beta x)^{2}+|d_{{\rm typical}}|^{2}}$.
$U_{0}$ is the transformation matrix to diagonalize $X_{p+1}$,
whose explicit form we present here for a later use, as
\beqa
\begin{array}{cc}
U_{0} &=\frac{1}{\sqrt{2 y }|d|_{{\rm typical}}}
\left(
\begin{array}{cc}
\sqrt{y+\beta x}|d|_{{\rm typical}}\ & 
\sqrt{y-\beta x}d^{*}_{{\rm typical}} \\
-\sqrt{y-\beta x}d_{{\rm typical}} & 
\sqrt{y+\beta x}|d|_{{\rm typical}}
\end{array}
\right). 
\end{array}\label{u0}
\eeqa  
Since it holds  $|d_{{\rm typical}}|^{2}=
|\tilde{c}(x_{\mu})_{{\rm typical}}|^{2}/2
\cong |\tilde{c}(x_{\mu})|^{2}_{{\rm typical}}/2$\footnote{
In the last equality we use the approximation
$|\tilde{c}_{0k}^{*}(t)_{{\rm typical}}
\tilde{c}_{0k'}(t)_{{\rm typical}}
u_{k}^{*}u_{k'}|\cong
(2\pi)^{p-1} \delta^{(p-1)}(k-k') (\Delta_{k})^{2}/2$
where $u_{k}=e^{ik_{\bar{\mu}}x^{\bar{\mu}}}$ and 
$\Delta_{k}$ is given in (\ref{deltak1}). 
This holds good because only the contributions of
$c_{k}$ and $c_{k'}$ with $k=k'$ has a coherent  phase factor 
while the others do not.} 
which we have obtained in the previous section,
the formula for the shape of one of the recombined branes is
\beqa
y(x,t) =\sqrt{(\beta x)^{2} 
+ \frac{A(t)^{2}}{2}\sqrt{\frac{2\beta}{\pi}}e^{-2\beta x^{2}} }.
\label{braneshape1}
\eeqa
This is the same form as obtained for D-strings in ref.\cite{hn}.
However, we can make a precise analysis on the shape of the branes at 
arbitrary $t$ based on (\ref{braneshape1}), 
since we have the explicit function of $A(t)$ for each case
including the overall factor and fine coefficients; 
We can know explicitly 
when and which applicable condition of the approximations 
(SYM, WKB and some others) breaks.
We will investigate the behavior of the system in detail using
(\ref{braneshape1}) with $A(t)$'s obtained in the previous section.

For the case (II),
its essential difference from the case (I) is that 
there are two scalar fields which have non-trivial 
VEV's of  $X_{p+1}$ and $X_{9}$.
Since the two do not commute each other
after tachyon condensation,
it is probable that one cannot determine 
definite positions of the branes in both of the
dimensions $x_{p+1}$ and $x_{9}$ simultaneously (i.e. uncertainty relation).
Here we assume that we can interpret each of $X_{p+1}$'s 
diagonal elements 
as the position of the two branes in the $x_{p+1}$  direction
if we focus only on the positions in $x_{p+1}$ direction
regardless of  $x_{9}^{(0)}$ direction.
We will return to this problem in the final section.

At the initial stage, the shape of each brane is approximately 
one part of hyperbola as in Fig. 5, 
which is consistent with our intuition; 
though $|d|_{{\rm typical}}$ is dependent on $x$ 
through the Gaussian $f_{0}(x)$,
its width $1/\sqrt{\beta}$ is broad and 
its differential coefficient is so small.
\begin{figure}[h]
\begin{center}
\includegraphics[width=8cm]{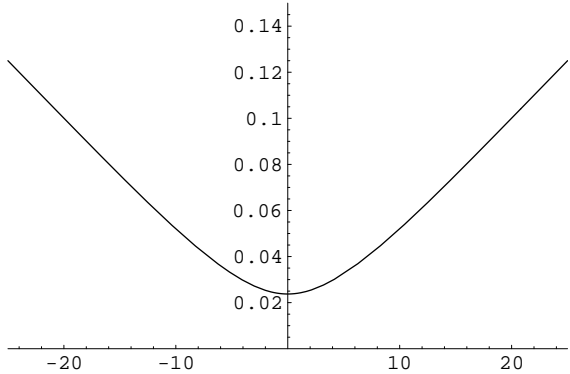}
\caption{
The shape of D-4-branes;
$\theta\simeq 2\beta=0.01$ and $A(t)=0.01$
}
\end{center}
\end{figure}
The larger $|d|_{{\rm typical}}$ becomes, 
the further the distance of the two D-brane becomes.
That is, $|d|_{{\rm typical}}$
corresponds directly to such a  
relative motion between the recombined branes as discussed 
in ref.\cite{hn}. 

So far, the shape of the p-branes is within the reach of our
intuition. 
After more time passes, however, a seemingly queer behavior
occurs; the shape of each brane deviates from the approximate
hyperbola and comes to have three extremes 
due to localization of tachyon condensation. 
This happens for the cases of D-p-brane for $p\ge 2$,
and is also expected to happen for the case of D-strings,
as we will show below.\footnote{
The appearance of such a shape was already discussed,  
but was {\it denied}
in ref.\cite{hn} in the case of D-strings.}

The values of $x$ giving extremes of $y(x,t)$ 
are formally given by
\beqa
x=0,\pm\sqrt{\frac{1}{4\beta}\ln \frac{2 A^4}{\pi\beta}},
\eeqa 
which means that each brane's shape has three extremes
if it holds
\beqa
A>(\pi\beta/2)^{1/4}\equiv A_{{\rm critical}}.\label{multi1}
\eeqa
Thus, all we have to do is to show that the condition 
(\ref{multi1}) is compatible with applicable conditions of 
the approximations we have used.
One  of the important conditions is  
that of the WKB approximation that
higher order terms of the fluctuations in the action 
do not disturb the behavior.
Since only $c$ and $d$ include the tachyonic field,
we estimate the orders of their second order terms $S_{2}$ and 
fourth order terms $S_{4}$ given respectively as
\beqa
S_{2}&\equiv& -T_{p}\int d^{p+1}x 2 i \beta (c^{*}d - c d^{*})\nonumber\\
S_{4}&\equiv& -T_{p}\int d^{p+1}x |c^{*}d-cd^{*}|^{2}\label{ss4}
\eeqa
and show $|S_{4}|\ll |S_{2}|$ below.

For D-p-branes with $p\ge 2$,
if we include the non-linear terms in the equation of motion, 
(\ref{expansion2}) (with $\tilde{c}_{0,k}$ replaced 
by its time-dependent typical value)
are not the solutions to the equation of motion any more. However, 
since each of $f_{n}$ and $u_{k}$
forms a complete set,
the expansion is still available
as a type of decomposition of d.o.f..
So, we substitute (\ref{expansion2}) for $S_{2}$ and $S_{4}$
and discuss their amplitudes. 
Neglecting the contribution of non-tachyonic 
(i.e. non-blowing-up) modes,
we have 
\beqa
S_{2}&=&-T_{p}\int dt (-2\beta)A(t)^{2}
\nonumber\\
S_{4}&=&-T_{p}\int dt \sqrt{\frac{\beta}{\pi}} 
\int\frac{d^{p-1}{\bf k_{1}}}{(2\pi)^{p-1}}
\int\frac{d^{p-1}{\bf k_{2}}}{(2\pi)^{p-1}}
\int\frac{d^{p-1}{\bf k_{3}}}{(2\pi)^{p-1}}
\tilde{c}_{0,k_{1}}^{*}\tilde{c}_{0,k_{2}}\tilde{c}_{0,k_{3}}^{*}
\tilde{c}_{0,-k_{1}-k_{2}-k_{3}}.\label{s4}
\eeqa
since $A(t)^{2}\equiv \int d^{p-1}{\bf k}/(2\pi)^{p-1} 
<|\tilde{c}_{0,k}|^{2}>$.
The factor $\sqrt{\beta/\pi}$ comes from extra 
normalization constants of $f_{0}$'s. 
To carry out the integrations with respect to momenta in (\ref{s4}) 
is difficult, so we make order estimation of $S_{4}$,
which will later prove to suffice here;
first we estimate a typical value of $k=\sqrt{k_{\bar{\mu}}^{2}}$
in condensation of $\tilde{c}(x_{\mu})$ using 
$k_{{\rm av}}=\sqrt{\bar{k^{2}}}$ defined at the last part of section 2;
it is about $30 \sim 5$ percent of $\sqrt{k^{2}_{{\rm
max}}}=\sqrt{2\beta}$ and
monotonically decreasing during $t\ge 10 T^{(I)}$ in the case (I)
or $t\ge 10 T^{(II)}$ in the case (II).
Next, we choose some appropriate value of $k_{{\rm av}}$ and 
approximate $\tilde{c}_{0,k_{i}}$ for $i=1,2,3 $ in (\ref{s4})  
to $|\tilde{c}_{0,k_{av}}|$,
defining $A_{{\rm av}}(t)^{2}\equiv \int d^{p-1}{\bf k}/(2\pi)^{p-1} 
<|\tilde{c}_{0,k_{{\rm av}}}|^{2}>$. 
Then, we can show that the behavior of $A(t)_{{\rm av}}$ is 
not so different from that of $A(t)$, and 
we estimate  $S_{4}$ as
\beqa
S_{4}\sim-T_{p}\int dt  
\frac{\Omega_{p-2}(2\beta)^{p/2}}{(2\pi)^{p-1/2}}
(A_{{\rm av}}(t))^{4}.
\eeqa 
Thus, the condition for $|S_{4}|\ll |S_{2}|$ is
\beqa
A(t)\ll \sqrt{\frac{(2\pi)^{p-1/2}}
{\Omega_{p-2}(2\beta)^{(p-2)/2} }}.
\label{wkbdp}
\eeqa
We can easily see that for a small angle $\theta\simeq 2\beta$ and $p\ge 2$,  
the inequality (\ref{wkbdp}) is sufficiently
compatible with (\ref{multi1}).
The other conditions, especially that of SYM that the displacement of
branes should be 
smaller than $l_{s}$ ($A(t)< l_{s}$) and D-branes should have 
some low velocity ($\dot{y}\ll 1(=c)$ at all the value of $x$),
are also shown easily to be satisfied.  
Therefore, the shape of the recombined D-p-brane for $p\ge 2$ 
surely come to have three extremes.
An example is drown in Fig.6.
This happens essentially due to the localization of tachyons 
around $x$,
since the factor $e^{-2\beta x^{2}}$ in (\ref{braneshape1}) is directly
responsible for the multiple extremes.
\begin{figure}[h] 
\begin{center}
\includegraphics[width=8cm]{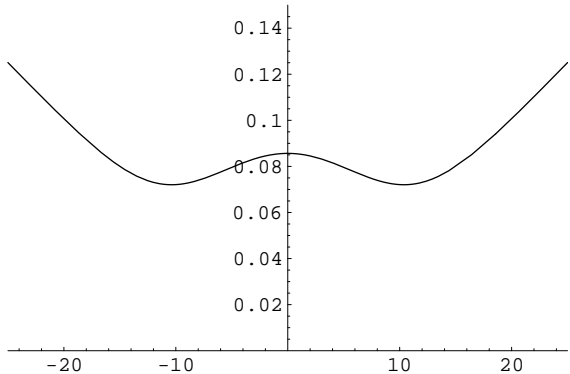}
\caption{
The shape of a D-brane with multiple extremes;
$2\beta=0.01$ and $A(t)=0.13$
}
\end{center}
\end{figure}

For the case of D-strings,
the case is a bit more subtle, but we argue that 
the same phenomenon is expected to happen. 
In this case 
$S_{2}$  and $S_{4}$ are written as
\beqa
S_{2}&=&-T_{D1}\int dt (-2\beta)|\tilde{c}_{0}(t)|^{2}_{{\rm typical}}
\nonumber\\
S_{4}&=&-T_{D1}\int dt \sqrt{\frac{\beta}{\pi}}
|\tilde{c}_{0}(t)|^{4}_{{\rm typical}}.
\label{s41}
\eeqa
If we substitute $\tilde{c}_{0}(t)=A(t)$
for the condition $|S_{4}|< |S_{2}|$, we have
\beqa
A(t) < (4\pi \beta)^{1/4}.\label{wkbcond2}
\eeqa
If one consider the equation of motion for $\tilde{c}_{0}(t)$,
\beqa
\frac{d^{2} \tilde{c}_{0}(t)}{dt^{2}}= 2\beta \tilde{c}_{0}(t) -2
|\tilde{c}_{0}(t)|^{2}\tilde{c}_{0}(t),\label{eomhi}
\eeqa
(\ref{wkbcond2}) is replaced by $A(t) < (\pi \beta)^{1/4}$.
Anyway, there is, a little though , a range 
for $A(t)$ to be allowed to have three extremes
within second order approximation: 
$(\pi \beta/2)^{1/4}< A(t) < (\pi \beta)^{1/4}$.
This may seem subtle, but if we solve numerically the 
equation of motion for $\tilde{c}_{0}(t)$
we can see that $|\tilde{c}_{0}(t)|$ really develop to exceed 
$A_{{\rm critical}}$ in (\ref{multi1}).
The other conditions of the approximations are again
shown to be satisfied. 
Thus, it is probable that
the shape of the D-string also comes to have three 
extremes.

The physical interpretation of this seemingly queer behavior  is 
as follows;
the energy released via tachyon condensation 
pushes the recombined branes away from each other, but
it is given only to the local part of the branes around $x=0$
due to localization of tachyon condensation, so, only 
the part is much accelerated.
Though the D-branes have a large tension, 
they also have a large inertia, and 
when the given energy of the local part is large enough,
the branes extend, surpassing the tension, to form three extremes.
That is, {\it localization of tachyon condensation or that of  
the released energy, and the (large) inertia causes the shapes of branes 
to have three extremes.}

The recombined picture has been obtained, and that the seemingly
queer behavior can also be explained from 
physical point of view. Thus, 
we can say that 
each diagonal element of $X_{p+1}$ describes the shape of each D-brane,
and that
in the gauge where 
$X_{p+1}$ is diagonalized (which we denote as ``physical gauge''),
the correspondence of degrees of freedom 
is direct between open strings and
the matrices $X_{i}$ and $A_{\mu}$;
for example, (1,1) component of $A_{\mu}$ corresponds to excitations of 
open strings whose both ends are on a first D-brane, etc..  

\section{Creation of open string pairs} 
\setcounter{equation}{0}
\setcounter{footnote}{0}

Let us next discuss creation of open string pairs.
We mainly concentrate on case (I) in this section:
We first discuss the case of D-strings, and then
that of D-p-branes with $p \ge 2$.

The blow-up of the gauge field appears in the off-diagonal elements of 
$A_{p}$ (in this case of course $p=1$) and a typical value of 
its electric flux is
\beqa
\begin{array}{ccc}
F_{0p}(x_{\mu})_{{\rm typical}}&
=\left(
\begin{array}{cc}
0 &  \partial_{0} c^{*}_{{\rm typical}}\\
\partial_{0} c_{{\rm typical}} & 0
\end{array}
\right) 
&\cong \frac{1}{\sqrt{2}}
\left(
\begin{array}{cc}
0 &  \dot{A}(t)f_{0}(x)e^{-i\delta}\\
\dot{A}(t)f_{0}(x)e^{i\delta} & 0
\end{array}
\right) 
\end{array}
\eeqa
where a dot represents a time-derivative and $\delta$ is a constant
phase of $c_{{\rm typical}}$ (and $\tilde{c}_{{\rm typical}}$).
This flux has a non-vanishing contribution to the energy density 
${\cal H}_{{\rm ele}}(x)$ as
\beqa
{\cal H}_{{\rm ele}}(x)=\frac{1}{2g_{s}}(\dot{A})^{2}(f_{0})^{2},
\eeqa
so this excitation should have some physical meaning.
What is the physical meaning of the blow-up? 
We argue that it means that {\it open string pairs (each composed of 
two strings with opposite orientations) connecting the 
two recombined D-branes  are created}.
We will explain the reason in the following some paragraphs.

First, we show that this blow-up correspond to 
excitations of open strings connecting the two branes.
Here we would like to confirm some basic facts about SYM as low energy
effective theory (LEET) of D-branes: 
As discussed in \cite{witten}, each D-brane  (${\rm Dp}_{I}$ for I=1,2) 
has U(1) symmetry
(${\rm U(1)}_{I}$), and open strings starting at ${\rm Dp}_{1}$ and 
ending at ${\rm Dp}_{2}$ (which we denote as 1$\to$2 strings) 
have ${\rm U(1)}_{1}\times {\rm U(1)}_{2}$ charges (-1,1), and 
2$\to$1 strings have charges (1,-1). 
In the U(2) SYM plus adjoint Higgs theory as LEET of branes, 
effective fields corresponding to the strings also have such charges, and
if degrees of freedom (d.o.f.) of ${\rm Dp}_{I}$ 
correspond to (I,I)-components of
the U(2) matrices for I=1,2 (as in the case of the physical gauge),
one can represent
${\rm U(1)}_{1}$ and ${\rm U(1)}_{2}$ transformation as 
$U_{1}={\rm diag}(e^{i\alpha_{1}},1)$ and $U_{2}={\rm diag.}
(1,e^{i\alpha_{2}})$ 
for parameters $\alpha_{1}$ and $\alpha_{2}$.
Suppose we use a unified symbol of the U(2) field $\hat{A}_{M}$, 
which is equal to 
$A_{\mu}$ for $M=\mu$ (=$0,1,\cdots,p$) and is equal to  
$X_{i}$ for $M=i$ (=$p+1,\cdots,9$). 
Then, $\hat{A}_{M}$ transforms under  
the ${\rm U(1)}_{1}$ and ${\rm U(1)}_{2}$ with constant 
$\alpha_{I}$'s as
\beqa
\begin{array}{cc}
\hat{A}_{M}
=\left(
\begin{array}{cc}
\hat{a}_{M} &  \hat{b}_{M} \\
\hat{c}_{M}  & \hat{d}_{M} 
\end{array}
\right) 
&\to  \hat{A}_{M}^{'}=
\left(
\begin{array}{cc}
\hat{a}_{M} &  \hat{b}_{M}e^{i(\alpha_{1}-\alpha_{2})} \\
\hat{c}_{M} e^{-i(\alpha_{1}-\alpha_{2})} & \hat{d}_{M}
\end{array}
\right),
\end{array}\label{am1}
\eeqa
so, d.o.f. of $\hat{c}_{M}$ correspond to those of $1\to 2$ strings and 
d.o.f. of $\hat{b}_{M}$ to those of  $2\to 1$ strings.

In our case it is when $X_{p+1}$ is diagonalized 
(i.e. in the physical gauge) that 
d.o.f. of ${\rm Dp}_{I}$ correspond to (I,I)-components of
the U(2), so we have to see the gauge field in the physical gauge.
Using $U_{0}$ in (\ref{u0}) (and abbreviating $c_{{\rm typical}}$ to $c$
here),  we have the field in the gauge 
$\hat{A}_{M}^{'}=U_{0}\hat{A}_{M}U_{0}^{-1}
+i U_{0}\partial_{M}U_{0}^{-1}$ as
\beqa
\begin{array}{ccc}
A_{0}^{'} & =&\frac{\beta x}{4y^{2}}
\frac{\partial_{0}c^{*} 
 c 
+c^{*} 
\partial_{0}c} 
 {|c
|^{2}}
\left(
\begin{array}{cc}
0 &c^{*}
\\
 c
& 0
\end{array} 
\right)\\
A_{p}^{'} &
=&\{1-\frac{\beta+2(\beta x)^{2}}{2y^{2} }\}
\left(
\begin{array}{cc}
0 &c^{*}
\\
c
& 0
\end{array} 
\right)\\
X_{p+1}^{'}& =&
\left(
\begin{array}{cc}
y & 0\\
0 & -y
\end{array}
\right),
\end{array}
\label{newgaugeff}
\eeqa
and the electric flux as
\beqa
\begin{array}{ccc}
F_{0p}^{'}(x_{\mu})
&
=U_{0}F_{0p}(x_{\mu})
U_{0}^{-1}=
&\left(
\begin{array}{cc}
0 & \partial_{0} c^{*}\\
\partial_{0} c & 0
\end{array}
\right).\label{eflux1} 
\end{array}
\eeqa
That is, the flux is invariant under the transformation via $U_{0}$; 
the blow-up arises only in its off-diagonal parts and hence, 
according to the above basic facts, 
the blow-up of the electric flux is due to excitations
of 1$\to$2 and 2$\to$1 strings.   

Then, how is the ratio of 1$\to$2 and 2$\to$1 strings?
We can see that via charge densities $j_{0}^{(I)}$ (for I=1,2)
of ${\rm U(1)}_{I}$ current $j_{\mu}^{(I)}$;
the above basic facts means that 
$j_{0}^{(1)}$ measures 
the number density of endpoints of open strings incoming to 
${\rm Dp}_{1}$
minus that of strings outgoing from ${\rm Dp}_{1}$
(if it is appropriately normalized), and $j_{0}^{(2)}$ is equal to
$-j_{0}^{(1)}$ in this setting. 
In the notation of $\hat{A}_{M}$ in (\ref{am1}), 
$j_{0}^{(1)}$ is written via Noether's theorem as
\beqa
j_{0}^{(1)}(=-j_{0}^{(2)})=&i\{ 
(\hat{b}_{M}\partial_{0}\hat{c}_{M}-\hat{c}_{M}\partial_{0}\hat{b}_{M})
-(\hat{b}_{M}\partial_{M}\hat{c}_{0}-\hat{c}_{M}\partial_{M}\hat{b}_{0})\}\nonumber\\
&-2(\hat{a}_{0}-\hat{d}_{0})\hat{b}_{M}\hat{c}_{M}
+(\hat{a}_{M}-\hat{d}_{M})(\hat{b}_{0}\hat{c}_{M}
+\hat{b}_{M}\hat{c}_{0}).\label{j0}
\eeqa
Substituting the solution (\ref{newgaugeff}) for (\ref{j0}),
$j_{0}^{(1)}$ is shown to vanish, which means that
the number density of endpoints of open strings
incoming to ${\rm Dp}_{1}$ at the point $x$ is equal to that
of strings outgoing from ${\rm Dp}_{1}$ at $x$.
This is also the case with $j_{0}^{(2)}$ and ${\rm Dp}_{2}$.
That is, open string pairs connecting the two decaying D-p-branes 
are created, each pair of which is composed of 1$\to$2 and 2$\to $1
strings! 
\footnote{We 
can also look into the configuration of fundamental strings in our solution 
(\ref{newgaugeff}) via electric source current $J_{0M}(x_{N})$ of
spacetime NSNS field $B_{MN}$ by
using the results of ref.\cite{raams}, as done in ref.\cite{khw}.  
Then, we can show that $J_{0M}$ vanishes in this case, which is 
consistent with $j_{0}^{(1)}=j_{0}^{(2)}=0$ and our interpretation 
of it.}

Furthermore, excitations of the scalar and gauge fields in
(\ref{newgaugeff}) also
includes other 
components which correspond to such open string pairs.\footnote{
I thank K.Hashimoto for letting me notice these extra contributions.} 
We can extract
the components by 
writing all the excitations
in terms of U(2) covariant forms in the physical gauge, as
\beqa
D'_{0}X_{p+1}'&=& U_{0} D_{0}X_{p+1}U_{0}^{-1}
=\left(
\begin{array}{cc}
\partial_{0}y & \beta x \partial_{0}d^{*}/y \\
\beta x \partial_{0} d/y & - \partial_{0}y
\end{array}\right),
\nonumber\\
D'_{p}X_{p+1}'&=&U_{0} D_{p}X_{p+1}U_{0}^{-1}=
\left(
\begin{array}{cc}
\partial_{x}y & -(\beta -2|d|^{2})d^{*}/y \\
-(\beta -2|d|^{2})d/y & -\partial_{x}y
\end{array}
\right).\label{dx1}
\eeqa
By T-dualizing with respect to $x_{p+1}$ direction, 
$D'_{0}X_{p+1}'$ and $D'_{p}X_{p+1}'$ correspond respectively to
the electric and magnetic fluxes $F'_{0,p+1}$ and $F'_{p,p+1}$. 
Thus, based on the same reason as 
the case of the electric flux $F'_{0p}$,
we argue that only diagonal elements in (\ref{dx1}) correspond
to d.o.f. of D-1-branes and the remaining
off-diagonal components in (\ref{dx1}) correspond
to d.o.f. of open strings connecting the branes. 
Consistency check of this argument is given in the following way:
Non-vanishing part of the total energy density of th system is 
\beqa
{\cal H}_{{\rm total}}(x)=\frac{T_{D1}}{2}Tr\{1_{2\times 
2}+\frac{1}{2}(D'_{p}X_{p+1}')^{2}+\frac{1}{2}(D'_{0}X_{p+1}')^{2}
+\frac{1}{2}(F'_{0p})^{2}\},\label{etotal}
\eeqa
and contribution of the diagonal elements of (\ref{dx1}) to 
${\cal H}_{{\rm total}}(x)$ is 
\beqa
{\cal H}_{{\rm diag}}(x)=T_{{\rm D1}}\{2+2\cdot
\frac{1}{2}(\partial_{x}y)^{2}
+2\cdot \frac{1}{2} (\partial_{0}y)^{2}
+0\}.\label{ediag}
\eeqa
while that of the off-diagonal elements of (\ref{dx1}) and $F'_{0p}$
is
\beqa
{\cal H}_{{\rm off-diag}}=T_{{\rm D1}}\{
(\beta-2|d|^{2})^{2}|d|^{2}/y^{2}
+(\beta x)^{2}|\partial_{0}d|^{2}/y^{2}+|\partial_{0}c|^{2}\}.
\label{eoffdiag}
\eeqa
The first and second term in (\ref{ediag}) are rewritten within second
order approximation 
as $2\sqrt{1+(\partial_{x}y)^{2}}$, the energy density due to the
D-strings' existence 
(tension), and the third term
is of course the density of the D-strings' kinetic energy. That is, only
${\cal H}_{{\rm 
diag}}(x)$ 
just gives the energy density that the D-strings themselves have.    
Thus, it is plausible that the remaining energy 
${\cal H}_{{\rm off-diag}}$ is the energy due to 
interaction of D-branes, which is consistent with our argument that
open string pairs connecting the D-branes have the rest of 
the energy.\footnote{Further confirmation of our argument
that open string pair creation are created between the D-branes
will be given in ref.\cite{hashi}.}

We can also estimate roughly 
the number density of the created open strings;
since SYM describes the behavior of low energy modes or almost zero
modes of open strings, the strings connecting the two D-branes
should be nearly straight. 
Since the tension, or 
the energy density of fundamental strings 
is constant (=1 in our notation), the number density of 
the open strings (per unit length) 
$n(x,t)$ is equivalent to
the energy density divided by the displacement of the two D-p-branes 
in $x_{p+1}$ direction.\footnote{In fact, open strings have momenta
since the off-diagonal elements to momentum density is non-vanishing. 
Here, we ignore the contribution of momenta as an approximation
and and estimate the number density.} 
Thus, it is obtained as 
\beqa
n(x,t)=\frac{{\cal H}_{{\rm off-diag}}(x,t)}{2y(x,t)}
=\frac{|d|^{2}}{2 y^{3}}\{(\beta-2|d|^{2})^{2}
+2\cdot 2\beta(\beta x)^{2}+2\beta |d|^{2}\}
\label{numb1}
\eeqa
where we have used $\partial_{0}d =\sqrt{2\beta}d$
and  $\partial_{0}c =\sqrt{2\beta}c$.
A distribution of $n(x,t)$ for $g_{s}=10^{-4}$ and 
$A(t)=0.1$ is given in Fig.7,
and the situation is roughly depicted in Fig.8.
\begin{figure}[h]
\begin{center}
\includegraphics[width=8cm]{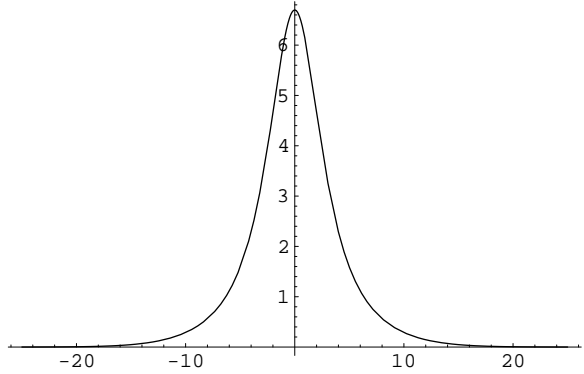}\\
\caption{Number density of open strings for $A(t)=0.1$ with $g_{s}=10^{-4}$}
\end{center}
\end {figure}
\begin{figure}[h]
\begin{center}
\includegraphics[width=12cm]{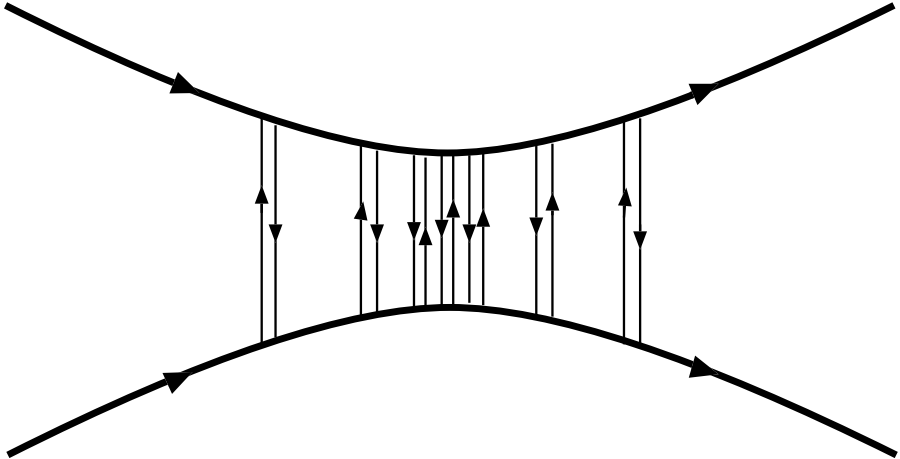}\\
\caption{A rough picture how open string pairs are created}
\end{center}
\end{figure}

In fact, a total number of the fundamental string in Fig.7 is
$N\equiv \int dx n\cong 50.14$, i.e. not an even integer. 
This is due to our partially incomplete analysis,
in which we do not consider the effect of quantization of charges 
or fluxes. Actually, $n(x,t)$ is an
expectation value 
or a probability of distribution of number density 
for open string to be created, and the real process
should happen in a quantum theoretic way; each pair will be created
rather ``suddenly''  with the above probability.

Before moving on to the case of D-p-branes with $p\ge 2$,   
we discuss distribution of the released energy.
The energy released by the shortening of D-branes is 
\beqa
E_{{\rm released}}&=&T_{D1}\int dx \  2
\sqrt{1+(\partial_{x}y)^{2}}|_{t=0}
-T_{D1}\int dx \ 2 \sqrt{1+(\partial_{x}y)^{2}}|_{t}\nonumber\\
&\simeq & T_{D1}\int dx \{\beta^{2}-(\beta-2|d|^{2})^{2}(\beta x)^{2}
/y^{2} \},
\label{ereleased}
\eeqa
and it is used partly to accelerate
D-branes and partly to create the open string pairs.
Then, how much is the ratio of the energy distribution?
D-branes' kinetic energy $E_{D1\ {\rm kinetic}}$ and the energy used 
to create open string pairs $E_{{\rm open} }$ are given respectively 
as
\beqa
E_{D1\ {\rm kinetic}}=T_{D1}\int dx
(\partial_{0}y)^{2}\label{ekinetic}, \ \ 
E_{{\rm open} } =T_{D1} \int dx 
{\cal H}_{{\rm off-diag}}
\label{eopen}
\eeqa
(Of course it holds $E_{{\rm released}}=E_{D1\ {\rm kinetic}}
+E_{{\rm open}}$.)
Since we know all the quantities in the formulas of the energies,
we can calculate the ratios    
$E_{D1\ {\rm kinetic}}/E_{{\rm released}}$ and  
$E_{{\rm open} }/E_{{\rm released}}$ at the time $t$ numerically.
The result is 
\beqa
\left\{
\begin{array}{ccc}
E_{D1\ {\rm kinetic}}/E_{{\rm released}}&< 10 \% &\nonumber\\
E_{{\rm open} }/E_{{\rm released}}&> 90 \% & {\rm for}\  A(t)<0.1 
\end{array}\right.
\eeqa
as in Fig.9. and  Fig.10.
\begin{figure}[h]
\begin{center}
\includegraphics[width=8cm]{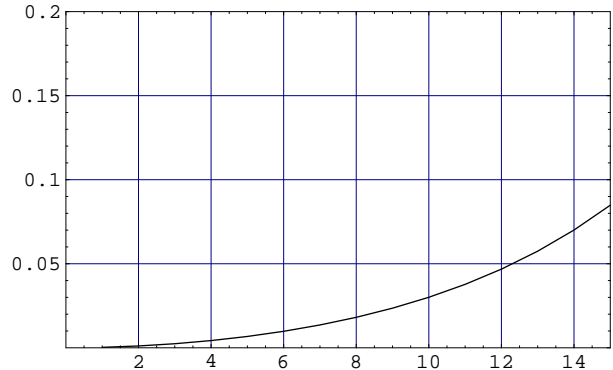}
\caption{$E_{D1\ {\rm kinetic}}/E_{{\rm released}}$ vs. $t(\times l_{s})$}
\end{center}
\end{figure}
\begin{figure}[h]
\begin{center}
\includegraphics[width=8cm]{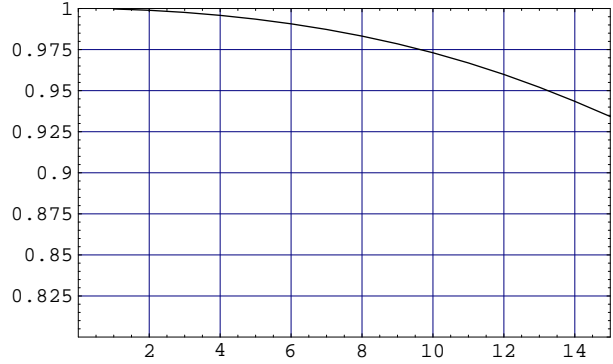}
\caption{$E_{{\rm open} }/E_{{\rm released}}$ vs. $t(\times l_{s})$}
\end{center}
\end{figure}

\noindent
That is, {\it most part of the released energy is used to create the open
string pairs connecting 
the branes}; only a little part of it is used to accelerate the D-branes.
Thus, from the viewpoint of D-branes' motion,
dissipation of the energy of the branes is very large at least
at the initial stage of decay. We note that the order of  $E_{{\rm open}}$
in terms of $g_{s}$ is ${\cal O} ((g_{s})^{0})$, and asymptotic behavior
of $E_{{\rm open}}$ is $E_{{\rm open}}\propto 
e^{2\sqrt{2 \beta}t}$ or $e^{\sqrt{2 \beta}t}$ (its exponent is
 decreasing along with time) for $A(t)<0.1$.

Next we study the case of D-p-branes with $p\ge 2$.
The main difference is that they 
have extra world-volume coordinates $x_{\bar{\mu}}$, and hence 
the condensing tachyon field $\tilde{c}$
(and $U_{0}$) can have dependence of $x_{\bar{\mu}}$.
Then, in addition to $A_{0}^{'}$, $A_{p}^{'}$ and $X'_{p+1}$  
in (\ref{newgaugeff}), 
$A_{\bar{\mu}}$ ($\bar{\mu}=1,\cdots,p-1$) arises through 
the transformation 
via $U_{0}$, and both U(2) electric and  
magnetic flux have non-vanishing contribution to the energy.
A naive form of $A_{\bar{\mu}}$ is a bit complicated, but 
using the approximation in the footnote under (\ref{u0}),
we have 
\beqa
A^{'}_{\bar{\mu}}\cong\frac{1}{2y}
\left(
\begin{array}{cc}
0 & \partial_{\bar{\mu}}c^{*}\\
\partial_{\bar{\mu}}c
& 0
\end{array}
\right).\label{amubar}
\eeqa
Non-vanishing fluxes and covariant derivatives of $X$'s are
\beqa
F_{\bar{\mu}p}^{'}(x_{\mu}) 
&=&\left(
\begin{array}{cc}
0 &  \partial_{\bar{\mu}}c^{*} 
\\
\partial_{\bar{\mu}}c 
& 0
\end{array}
\right)\\
D'_{\bar{\mu}}X'_{p+1}&=&
\left(\begin{array}{cc}
\partial_{\bar{\mu}}y & \beta x \partial_{\bar{\mu}}d^{*}/y\\
\beta x \partial_{\bar{\mu}}d/y & -\partial_{\bar{\mu}}y
\end{array}
\right),
\label{mflux1} 
\eeqa
in addition to (\ref{eflux1}) and (\ref{dx1}). 
It should be stressed  that the above is U(2) magnetic flux but 
not U(1)'s one on each of the branes.
Thus, even if we consider the case of D-3-branes, its existence does not 
mean that D-strings are created. 
Instead, according to the basic facts stated above, 
these d.o.f. again correspond to the d.o.f. of fundamental strings,
and $j_{0}$ is again shown to vanish by substituting (\ref{newgaugeff}) 
and (\ref{amubar}) (with $x_{\bar{\mu}}$ dependence 
taken into account). 
Therefore, we conclude that  
in the case of recombination of D-p-branes with $p\ge 2$,
open string pairs connecting the two D-branes are created 
while they are decaying.

Let us consider the number density of the string.
In the case with $p\ge 2$,
$|\partial_{0}d|^{2}$ and $|\partial_{0}c|^{2}$ in ${\cal H}_{{\rm off-diag}}$
(in (\ref{eoffdiag})) in the case of D-strings are replaced by 
$|\partial_{\hat{\mu}}d|$ and $|\partial_{\hat{\mu}}c|$ 
where $\hat{\mu}=0,\cdots,p-1$. But then, if we use $<<g(k)>>$ defined in
(\ref{gk1}), it holds 
$|\partial_{0}d|^{2}\simeq<<2\beta-k^{2}>>(f_{0})^{2}/2$ and 
$|\partial_{\bar{\mu}}d|^{2}\simeq <<k^{2}>>(f_{0})^{2}/2$, so
$|\partial_{\hat{\mu}}d|$ gives  $2\beta |d|^{2}$ and
the number density formula (\ref{numb1}) holds again,
though in this case it is a density in a p-dimensional space. 

The released energy $E_{r}(t)$ is
\beqa
E_{r}(t)=-T_{Dp}\int d^{p}x\ [2\sqrt{1+(\partial_{x}y)^{2}
+(\partial_{\bar{\mu}}y)^{2}}\ ]^{t}_{t=0}.
\eeqa
Since an extra minus term $-(\partial_{\bar{\mu}}y)^{2}|_{t}
=-<<k^{2}>>(f_{0})^{2}|d|^{2}/y^{2}$ is added
to $E_{r}(t)$, the density part of the released energy is a bit reduced
from that in the formula (\ref{ereleased}), but the density of energy
used to create open string pairs has the same form as that in
(\ref{numb1}). (The density of D-p-branes' kinetic energy is reduced in
 terms of $A(t)$.) Thus, in the case of D-p-branes with 
$p \ge 2$, {\it the ratio of the energy used to create open
 string pairs to the released energy is more than that in the case of
 D-strings}, and we can say again that most part of the released energy 
  is used to create the open string pairs.

\section{Creation of closed strings}
\setcounter{equation}{0}
\setcounter{footnote}{0}


Let us assume $g_{s}\ll 1$. Then, the branes can be regarded as 
heavy objects, and after recombination, 
they start to move away from each other, though  very slowly,
but {\it with some acceleration} in opposite directions. Thus, 
one can easily expect that the system has non-trivial time-evolution
behavior of mass quadrupole moments, which causes gravitational 
radiation. Below, going out of the framework of SYM,
we show that the above expectation is true, and 
evaluate its energy flux, based on the time-evolution of 
the D-branes' shapes. In addition, we discuss creation of
closed strings for configurations of D-branes beyond the reach of SYM. 
In this section we mainly discuss the case (I) again, 
but we expect that essential part of the 
discussion also holds in the case of the case (II).

Since the coupling of a D-brane to the spacetime metric is 
$G_{10}M\simeq {\cal O}(g_{s})$ (where $G_{10}=g_{s}^{2}/(M_{10})^{8}$ is 
the gravitational constant, $M_{10}$ is a 10-dimensional mass parameter
and $M$ is the brane's mass),   
the background spacetime can be regarded as a flat one and
the wave we will consider is a deviation around the flat spacetime. 
Since we are dealing with  a motion of D-branes with a relatively slow
velocity in order to apply SYM, a non-relativistic approximation 
is applicable. In addition, the motion of the D-branes are 
localized around $x=0$ at least at the initial stage, 
so the following formulae can be applied to this case 
(see e.g. \cite{gravi});
the lowest order radiation of gravitational wave is 
electric quadrupole radiation, and
the mass quadrupole moment of a collection of particles with 
masses $\{ m_{I} \}$ in 10 spacetime dimensions is 
\beqa
I_{ij}=\sum_{I}m_{A}[x_{i}^{I}x_{j}^{I}
-\frac{1}{9}\delta_{ij}(x^{I})^{2}]
\eeqa
where $x_{i}^{I}$'s ($i=1,\cdots,9 $) represent (space) positions of 
a particle labeled by I.
The angular distribution of the energy flux due to the radiation
is given by 
\beqa
\frac{d^{2}E_{{\rm gravi}}}{dtd\Omega_{8}}=\frac{G_{10}}{8\pi} 
<(\frac{d^{3}I_{ij}}{dt^{3}})^{2}
-2(n_{i}\frac{d^{3}I_{ij}}{dt^{3}})^{2}
+\frac{7}{8}(n_{i}n_{j}\frac{d^{3}I_{ij}}{dt^{3}})^{2}>
\label{angra}
\eeqa
where $n_{i}\equiv (\bar{x}_{i}-x_{i})/|\bar{x}_{i}-x_{i}|$ where
$\bar{x} $ is the position of the observer and $<\cdots >$ denotes an
average over several wavelength here. 
We note that in (\ref{angra}) $x_{i}$ in $I_{ij}$ 
is to be evaluated at the retarded time.
To apply the above formulae, we regard the branes as an assembly
of infinitesimal hyper-planes.  
Then, 
$I_{ij}$ in the case of branes is translated into 
\beqa
I_{ij}=T_{p}\sum_{I=1}^{2}\int d^{p}x\
\sqrt{1+(\partial_{x}x_{i}^{I})^{2}}
[x^{I}_{i}x^{I}_{j}-\frac{1}{9}\delta_{ij}(x^{I}_{k})^{2}]
\eeqa
where the index $I(=1,2)$ denotes each of the branes.

Let us first discuss the case of D-strings in the case (I), 
the simplest case,
where D-strings moves (decays) toward positive and negative 
directions of $x_{2}$.
The $I_{ij}$ which have non-trivial components are 
\beqa
I_{11}&=&\frac{2}{9g_{s}}\int dx \sqrt{1+(\partial_{x}y)^{2}}
(8 x^{2}-y^{2})
\nonumber\\
I_{22}&=&\frac{2}{9g_{s}}\int dx \sqrt{1+(\partial_{x}y)^{2}}
(8 y^{2}-x^{2})\\
I_{33}&=&I_{44}=\cdots =I_{99}=-\frac{2}{9g_{s}}\int dx
\sqrt{1+(\partial_{x}y)^{2}}(y^{2}+x^{2})\nonumber
\eeqa 
Here we also assume that the branes has no velocity in $x$ direction.
If we use 9-dimensional polar coordinate system as
$x_{2}=\cos\theta_{1},x_{1}=\sin\theta_{1}\cos\theta_{2},
x_{3}=\sin\theta_{1}\sin\theta_{2}\cos\theta_{3},\cdots,
x_{8}=\sin\theta_{1}\cdots \sin\theta_{7}\cos\theta_{8},
x_{9}=\sin\theta_{1}\cdots \sin\theta_{8}$,
we have 
\beqa
\frac{d^{2}E_{{\rm gravi}}}{dtd\Omega_{8}}&=&\frac{G_{10}}{8\pi}
\{\frac{7}{2}\sin^{4}\theta_{1}(\frac{d^{3}I_{1}}{dt^{3}})^{2}
+\frac{7}{2}(1-\sin^{2}\theta_{1}\cos^{2}\theta_{1})
(\frac{d^{3}I_{2}}{dt^{3}})^{2}\nonumber\\
& &+\sin^{2}\theta_{1}
(\cos^{2}\theta_{1}\cos^{2}\theta_{2}-\sin^{2}\theta_{2})
\frac{d^{3}I_{1}}{dt^{3}}\frac{d^{3}I_{2}}{dt^{3}}
\}\label{eneflux1}\\ 
\frac{d E_{{\rm gravi}}}{dt}&=&\frac{10 \pi G_{10}}{99}
((\frac{d^{3}I_{1}}{dt^{3}})^{2}+(\frac{d^{3}I_{2}}{dt^{3}})^{2}
-\frac{1}{4}\frac{d^{3}I_{1}}{dt^{3}}\frac{d^{3}I_{2}}{dt^{3}})
\label{eneflux2}
\eeqa
where $I_{1}\equiv \int dx \sqrt{1+(\partial_{x}y)^{2}}y^{2}/g_{s}$ and 
$I_{2}\equiv \int dx \sqrt{1+(\partial_{x}y)^{2}}x^{2}/g_{s}$.
Since $|\frac{d^{3}I_{1}}{dt^{3}}|\gg |\frac{d^{3}I_{2}}{dt^{3}}|$
(because accelerated most are the part around $x=0$ of the D-branes but
$\partial_{x}y(x,t) $ vanishes at $x=0$.),
the first terms in (\ref{eneflux1}) and  (\ref{eneflux2}) are dominant.
Using the concrete formula for $y(x,t)$ in (\ref{braneshape1}),
we have the energy flux as
\beqa
\frac{d E_{{\rm gravi}}}{dt}&\simeq&\frac{10 \pi G_{10}}{99g_{s}^{2}}
16 (2\beta)^{3} A(t)^{4}
\eeqa
where $A(t)\propto e^{\sqrt{2\beta} t}$ is used.
It is of course non-vanishing.  
Thus, the decaying D-strings certainly radiate closed strings as
gravitational wave.

In the cases of D-p-branes with $p\ge 2$ in the case (I), 
only the diagonal elements
of $I_{ii}$ have non-trivial values again 
due to the integration with respect to $x_{\mu}$. 
To carry out the computation of the energy flux of the radiation, 
it is necessary to regularize the 
world volume of the D-branes, and if we do it, 
we can easily show that the energy flux
is shown to be non-vanishing, though the final result depends on
the regularization volume.
Thus, the decaying D-p-branes with $p\ge 2$ also 
radiate gravitational wave.  

In addition, radiation of RR gauge field should also occur after 
recombination of D-branes; each of the D-branes in the initial
settings is a BPS brane, so each part of the branes should be 
a source of the flux of RR (p+1)-form gauge field. 
Thus, RR gauge wave should also be radiated if the branes are 
accelerated and have non-vanishing multi-derivatives of 
their quadrupole moment with respect to time,\footnote{
We note that the lowest order RR gauge radiation is also their 
electric quadrupole one; 
the two D-branes are regarded to have RR charges of the
{\it same} signature because if one set $\theta=0$, the force due to RR
gauge field is repulsive.}  
which is the case we are considering.
We expect that the energy flux of the RR gauge wave 
has about the same order as that of the gravitational wave.
  
Let us discuss the order of the energy given to the fundamental strings
in the case of D-strings.
In this paper we do not discuss the scale of $M_{10}$, and set
$M_{10}=M_{s}(=1/l_{s})$ for simplicity.
Then, since $G_{10}=g_{s}^{2}$ and $A(t)^{2}\sim {\cal O}(g_{s})$,
the flux $d E_{{\rm gravi}}/dt\sim {\cal O}(g_{s}^{2})$.
The energy flux of the RR gauge radiation is expected to be the same
order as  $d E_{{\rm gravi}}/dt$.
Taking into account that $E_{{\rm gravi}}(t)\propto A(t)^{4}$ and 
the asymptotic behavior of $E_{{\rm open}}$, we have
the ratio of the energy given to the massless closed strings to that
of open strings is 
\beqa
\frac{dE_{{\rm massless \ closed}}/dt}{dE_{{\rm open}}/dt}
\propto 
g_{s}^{2}e^{2\sqrt{2\beta} t} {\rm or\ } g_{s}^{2}e^{3\sqrt{2\beta} t}.
\label{ratio1}
\eeqa
Thus, the energy given to the massless closed string is very small
at the initial stage compared with that of open strings, but
the ratio increases exponentially with a typical time-scale
$T^{(I)}=1/\sqrt{2\beta}$

As for the case (II), the shape of the D-branes in all direction
may not be defined due to the non-commutativity of the scalar fields,
but it is not plausible that cancellation
of the radiation would happen.
So, gravitational radiation and that of RR gauge field are strongly 
expected to occur as creation of closed strings.


Next, we want to discuss the behavior of the system beyond SYM.
If the distance between the D-branes becomes larger than $l_{s}$,
it is difficult to imagine that the created open strings extend 
unlimitedly, because the string has its tension.
It also seems unnatural if one take into account increase of entropy.  
Then, what will happen? Since open strings are created {\it in pairs} 
it is expected that {\it each pair of open strings connecting the branes
are cut to pieces to form some closed strings and two open strings},
each of the latter of which has its both ends on each of the two D-branes,
as Fig.11. 
\begin{figure}[h]
\begin{center}
\includegraphics[width=12cm]{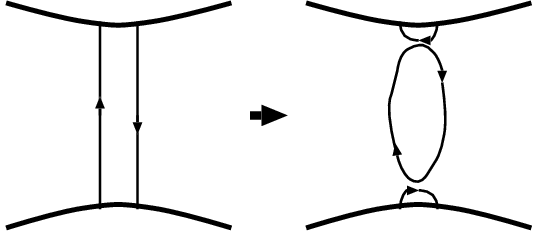}\\
\caption{}
\end{center}
\end{figure}
That is, such a picture seems to arise that the decaying (annihilating) 
D-branes leave many  closed strings 
behind (and radiate some of them) while producing many open strings 
which start from and end on each brane.
If this is true, the further the two branes go from each other,
the more possibility there arises to create closed strings. 
This expectation is consistent with tendency of the ratio of 
$dE_{{\rm closed}}/dt$ to $dE_{{\rm open}}/dt$.
The dominant mechanism to create closed strings in this setting might be  
the above way.
(Please note that if there was a single open string connecting 
the decaying D-branes, the energy of the system was not saved by creating 
a closed string.)

It is also to be noted that 
the above picture may be regarded as a generalization of 
Sen's conjecture for $D\bar{D}$ annihilation\cite{sen1}; 
in the case of  $D\bar{D}$ with no topological defects, 
the sum of the tensions of a D-brane and an anti-D-brane are 
precisely the height of the potential of the unstable vacuum and 
they vanish to be nothing.  
On the other hand, the  setting we are considering is a ``middle'' 
configuration of  
$D\bar{D}$  and parallel two BPS branes, the latter of which 
has definitely positive energy. Thus, our setting also has positive 
energy and hence there should remain something after the decay.
According to our speculation, it may be the closed and open strings 
stated above.

\section{Implications for an inflation model}
\setcounter{equation}{0}
\setcounter{footnote}{0}

In this section we consider implications of the above results
of the system  for the inflation model of the setting of two D-4-branes
at one angle in ref.\cite{oneangle}\cite{tye1}. 

First, we would like to speculate about the efficiency of the reheating
of the model.
According to the analysis in section 5, most part of the released energy via
tachyon condensation is used to create open string pairs connecting the
branes and only a little part is used to accelerate the branes,
at least at the initial stage of decay.
Thus, from the viewpoint of motion of the branes,
dissipation of the energy 
is rather large initially, and might be large afterward, so
the oscillation of the branes around some stable configuration 
might occur only a few times. 
That is, if this setting is applied to the inflation
scenario, it is possible that the reheating might be efficient.

Second, we would like to discuss the mechanism to create 
``reheated'' gauge fields and fermions;
from the viewpoint of cosmology they should be produced
at the end of the inflation, 
but the mechanism how are they produced in this setting of the inflation
model has not been clarified.
We speculate about a scenario of the mechanism based on our analysis:
Although we are considering a simpler setting model,
if a more complete model of brane-world is considered, 
the models should be the one in which at the end of inflation,
the configuration of branes are put so oriented
that they reproduce excitations of Standard Model
as given in e.g. ref.\cite{getjust}\cite{higgs} or some others.  
 (A prototype of such models are discussed  e.g. in ref.\cite{2angles}.) 
In the model, fields like fermions and gauge fields
arise as open strings connecting the branes or on each of the branes.
Our suggestion is that
the open string pairs connecting the branes like those in Fig.8
and the open strings on each of the branes like those in Fig.11
might directly correspond to the ``reheated'' fermions or gauge fields
created at the end of the inflation.
It would be interesting to find a framework describing the process in
the region of distance other than SYM
or beyond SYM, and discuss the above possibilities.

\section{Summary and discussion}
\setcounter{equation}{0}
\setcounter{footnote}{0}

We have investigated what happens in the process after the recombination 
of two D-p-branes at one angle for the two types of concrete 
initial conditions;
case (I):two D-p-branes have been parallel until $t=0$ but are put
intersected at one angle $\theta$ at the instant $t=0$, and 
case (II):one of parallel two
D-p-branes were rotated
by a small angle $\theta$ and 
are approaching each other very slowly.
In summary, the results are the following:
\begin{itemize} 
\item Explicit functions of typical amplitudes of condensing tachyonic
      fields (dependent on time and location) have been obtained. 
\item Time-evolution of the D-branes' shape after recombination
has been obtained, and seemingly queer behavior of the D-branes' shape that
each brane comes to have three extremes has been observed. Physical 
interpretation of it has also been given.
\item Pair-creation of (many) open strings with opposite orientations 
      connecting the two decaying branes has been shown to occur.
      Expectation value of number density of the created open strings
      has
been estimated,
      and energy distribution into the open strings has also been
      evaluated; it has been shown that most part of the energy released 
      via
      tachyon condensation is used to create the open string pairs, and
      dissipation of the D-branes' energy is large at the initial stage.
\item Creation of closed strings 
      as gravitational radiation from the branes
      has also been shown to occur; its energy flux has been evaluated
      quantitatively,
      In addition, a strong suggestion has been obtained that 
      closed strings are created out of open string pairs after the
      two branes get further from each other.
\item Speculations on implications of the above results for the  inflation
      model of the D-4-branes at one angle have been given: 
      dissipation of the branes' energy (due to the creation of the
      strings)
      might be large
      and reheating might be efficient. In addition, 
``reheated'' gauge fields and fermion which should appear at the end of the
      inflation, might correspond as the open strings created
      accompanying the decay of the D-branes.
\end{itemize}

Among the above, the most important result is 
the pair-creation of open strings connecting the decaying D-branes,
which strongly suggests that closed strings are created afterward.
These creations in the case of decay of D-p-branes at one angle
are in some senses like the case of decay of an unstable
D-brane (or s-branes\cite{stro1}) where rolling tachyons \cite{sen8}  
are used to discuss
production of open string pairs\cite{stro2}\cite{stro3}\cite{stro4} 
and closed strings 
\cite{closed1}\cite{closed11}\cite{closed2}\cite{closed3}
(see also \cite{st1}\cite{sen9} 
and other references therein).
Compared with the latter cases, the analysis used in this paper is
only perturbative and can follow the behavior of the system 
for only a rather short time-scale.
For further study of recombination of D-branes,
the approaches analogous to the case of an unstable D-brane
may be useful and is to be explored.

Finally, we discuss the problem of non-commutativity, 
uncertainty and diagonalization procedure (gauge choice) 
of VEV's of adjoint scalar fields 
$X_{p+1}$ and $X_{9}$, which appear in the solution of case (II).
The commutation relation of the two is
\beqa
[X_{p+1},X_{9}] = z(t) \left(
\begin{array}{cc}
0 & -d^{*} \\
d & 0 
\end{array}
\right). 
\label{comm}
\eeqa
It is plausible for the relation to give an uncertainty relation
\beqa
\Delta x_{p+1} \Delta x_{9} \ge |z(t)||dx,t|\propto 
|z_{0}-vt|A(t)e^{-\beta x^{2}}.
\label{uncer}
\eeqa
That is, this is a solution in which
the scale of uncertainty changes depending on location
and time.

Let us assume this uncertainty relation.
Then, the situation is as follows:
If one focuses on the positions in $x_{p+1}$ direction
and do not determine those in $x_{9}$ direction,
it seems possible to discuss the shapes of branes projected on 
$x_{p}x_{p+1}$-plane with some high accuracy. 
so, we have argued that the discussion in section 4 holds.
As for the ``shape'' (or profile) in  $x_{9}$ direction, 
suppose one diagonalizes $X_{p+1}$ and describes their positions
in the direction with the accuracy of the order $|d(x=0)|$,
i.e. the accuracy one can determine whether the branes are 
combined or not at $x=0$. 
Then, one is able to determine the positions in $x_{9}$ direction
only with the accuracy of order $|z|e^{-\beta x^{2}}$, 
which means that we cannot
tell whether the branes are separated or not at $x=0$, and
at the point a little away from $x=0$, one can tell the branes are 
separated but cannot determine the shape. And at the point more away from
$x=0$, one can tell the shape of the branes projected on 
$x_{p}x_{p+1}$-plane.

One may notice that there seems a bit problematic thing: 
When one takes the gauge which diagonalizes  $X_{9}$ instead,  
no evidence of recombination process can be observed.
It may be possible to attribute this problem to
the  gauge choice: On the one hand there is a gauge where 
geometric information of the branes is easy to extract 
(``physical gauge'' where $X_{p+1}$ is diagonalized).
On the other hand, 
there is a gauge where
the behavior of fluctuations like tachyon condensation
is easy to describe (the gauge $X_{9}$ is diagonalized).
 
In addition, we would like to speculate about the meaning of
diagonal elements of the scalar field which is not diagonalized; 
it might have some geometric meaning.
In the physical gauge, the VEV of $X_{9}$ is written as
\beqa
X_{9}^{'} = U_{0}X_{9}U_{0}^{-1} =
\frac{z(t)}{2\sqrt{(\beta x)^{2}+|d|^{2}}}
\left(
\begin{array}{cc}
 \beta x & -d^{*} \\
-d & -\beta x
\end{array}
\right). 
\label{x9}
\eeqa
The diagonal elements represent two kink-like shapes (profiles)
along x-direction with a width $|d|/\beta$ with 
both ends approaching $\pm z(t)/2$, which fits our expectation
for the branes' shapes projected on $x_{p}x_{9}$-plane.  
It may be a merely coincidence, 
but there might be a possibility that it has some geometric meaning.
Let us remember the solution of dielectric (spherical) 
D2/D0 bound states
given by Myers in ref.\cite{myers}. We concentrate
the N=2 (two D0) case. The solution is
\beqa
\begin{array}{ccc}
X_{1}= \frac{R}{\sqrt{3}}
\left(
\begin{array}{cc}
0 & 1 \\
1 & 0
\end{array}
\right),
 &
X_{2}= \frac{R}{\sqrt{3}}
\left(
\begin{array}{cc}
0 & -i \\
i & 0
\end{array}
\right),&
X_{3}= \frac{R}{\sqrt{3}}
\left(
\begin{array}{cc}
1 & 0 \\
0 & -1
\end{array}
\right)
\end{array}
\eeqa
where $ R$ is the extent of the spherical 
D-2-brane to which two D-0-branes are bound.
If one focuses on $x_{3}$ direction, 
the two D-0-branes are considered to locate in the north 
and south pole of the sphere in the $x_{3}$ direction. 
Then, one cannot determine the positions of two D0's
in $x_{1}$ and $x_{2}$ direction, but the zeros 
in diagonal elements of them may be interpreted as 
center of mass positions or expectation values of their positions.
Analogous to this case, 
it might be possible that 
the diagonal elements of $X_{9}^{'}$ describe center-of mass positions of
the branes in $x_{9}$ direction though the definite position is uncertain. 
The above point, related with non-commutativity, gauge
choice and uncertainty, 
would be interesting and to be explored in the future. 

\vskip 3ex
\noindent
{\large\bf Acknowledgement}
\vskip 3ex

I would especially like to thank Hidehiko Shimada for many fruitful
discussions, encouragement and much adequate advice 
in proceeding this work.
I also thank to Prof. K. Hashimoto and S. Nagaoka for useful 
communication and discussion.
I also like to thank Dr. Taro Tani for many useful discussion 
and encouragement, Prof. M. Kato for useful comments (questions),
directing my attention to some aspects I was not considering, 
and Y. Aisaka for teaching me how to use many computer tools.
I would like to thank Prof. H. Tye for giving me helpful
comments at the conference String Phenomenology 2003. 
Finally, I would like to thank members of Theoretical Physics Laboratory
at RIKEN for useful discussion.
This work is supported by Japanese Society for the Promotion of 
Science under Post-doctorial Research Program (No.13-05035).

\appendix

\vskip 3ex
\noindent
{\Large {\bf Appendix}}

\section{The width of a Gaussian wave function
for a mode with a time-dependent (negative) frequency-squared}
\setcounter{equation}{0}

In this appendix we first review derivation of the propagator (kernel) 
$K$ of the harmonic oscillator with a time-dependent frequency,
and then the fact that
an initially Gaussian wave function keeps to be
Gaussian within WKB approximation,
and finally present the formula for
the time-dependent width of the Gaussian in terms of
basis functions. See for more detail ref.\cite{fey}.

If one find two independent solutions $A(t)$ and $B(t)$ (which 
we denote as basis functions) to the ``WKB-approximated 
equation of motion
\beqa
\frac{d^{2} x(t)}{dt^{2}}= -\omega(t)^{2} x(t),
\label{eoma}
\eeqa
we can immediately write essential part of the propagator\cite{fey}
in terms of $A(t)$ and $B(t)$ up to WKB approximation.
The solution representing the  classical path of a motion 
from $(x_{0},t_{0})$ to $(x_{1},t_{1})$ is
\beqa
x(t)&=&\frac{1}{B_{1}A_{0}-A_{1}B_{0}}[\{ B_{1}A(t)-A_{1}B(t)\} x_{0}
+\{ -B_{0}A(t)+A_{0}B(t)\} x_{1}]\\
&\equiv & h_{0}(t) x_{0}+ h_{1}(t) x_{1}\label{classicalsol}
\eeqa
where we abbreviate $A(t_{1})$ as $A_{1}$, etc.
Substituting (\ref{classicalsol}) for the action 
$S=\int_{t_{0}}^{t_{1}} dt \ m(\dot{x}^{2}-\omega^{2}x^{2})/2$, 
and exponentiating 
the one times $i$, we have the propagator $K(x_{1},t_{1};x_{0},t_{0})=
F e^{iS_{{\rm cl}}}$ where $S_{{\rm cl}}$ is quadratic in $x$ as
\beqa
S_{{\rm cl}}=\frac{m}{2}\sum_{i=0,1}x_{i}x_{j} \alpha_{ij}(t_{1};t_{0})
\eeqa
where   
\beqa
\alpha_{ij}(t_{1};t_{0})\equiv \int_{t_{0}}^{t_{1}} dt 
[\frac{dh_{i}}{dt} \frac{dh_{j}}{dt} -\omega(t)^{2} h_{i}h_{j} ].
\eeqa
$F$ is a time-dependent normalization but not dependent on $x_{i}$,
which can be determined 
in several ways. 
(In this paper we determine the factor $F$ so that the propagator 
in each case $K$
satisfies the Schrodinger equation.)
Since the wave function for $t=t_{1}$ is given by
$\Psi(x_{1},t_{1})=\int dx_{0} K(x_{1},t_{1};x_{0},t_{0})
\Psi(x_{0},t_{0})$,
the absolute value of
an initially Gaussian wave function remains to be Gaussian,
since the above is only a Gaussian integral.
The behavior of the system  is represented by the 
time-dependent width of the Gaussian
obtained  in terms of the basis functions as
\beqa
\Delta(t)^{2}=\Delta(t_{0})^{2}
\frac{\alpha_{00}(t;t_{0})^{2}}{\alpha_{01}(t;t_{0})^{2}}
+\frac{1}{\Delta^{2}(t_{0})^{2}(\alpha_{01})^{2}}.\label{formaldelta}
\eeqa 
Thus, finding the basis functions leads directly to the 
time-evolution behavior of the width and the system up to WKB
approximation as far as the the initial wave function is Gaussian.
We note that normalizations of
$A(t)$ and $B(t)$ do not appear in
(\ref{classicalsol}) and hence also not in the final result.


\end {document}